\documentclass[aps,prl,reprint,superscriptaddress,nofootinbib,floatfix,raggedbottom]{revtex4-2}
\usepackage[utf8]{inputenc}
\usepackage[T2A]{fontenc}
\usepackage[russian,english]{babel}

\usepackage{amsmath,amssymb,bm}
\usepackage{booktabs}
\usepackage{xcolor}
\usepackage{hyperref}
\hypersetup{colorlinks=true,allcolors=blue}

\begin{document}

\title{Sub-percent accuracy in molecular electric quadrupole coupling: nuclear-size and relativistic effects in RaF}

\author{Leonid V.\ Skripnikov}
\email{skripnikov\_lv@pnpi.nrcki.ru,\\ leonidos239@gmail.com}
\homepage{http://www.qchem.pnpi.spb.ru}
\affiliation{Petersburg Nuclear Physics Institute named by B.P. Konstantinov of National Research Centre
``Kurchatov Institute'', Gatchina, Leningrad District 188300, Russia}
\affiliation{Saint Petersburg State University, 7/9 Universitetskaya nab., St. Petersburg, 199034 Russia}

\author{Michail Athanasakis-Kaklamanakis}
\altaffiliation[Present address: ]{JILA and University of Colorado, Boulder, Colorado 80309, USA}
\affiliation{KU Leuven, Instituut voor Kern- en Stralingsfysica, Leuven, Belgium}

\author{Carlos M.\ Fajardo-Zambrano}
\affiliation{KU Leuven, Instituut voor Kern- en Stralingsfysica, Leuven, Belgium}

\date{26.10.2026}

\begin{abstract}
We report a benchmark study of the electric-quadrupole hyperfine interaction in the electronic ground state of the RaF molecule, including finite-nuclear-size effects. The calculations include relativistic effects using the Dirac--Coulomb--Breit Hamiltonian and electron correlation up to perturbative quadruple excitations within the coupled cluster framework. At the achieved level of accuracy, penetration of the electronic density into the finite nuclear quadrupole distribution produces a non-negligible correction to the usual electric field gradient (EFG) description at the center of the nucleus and must be considered. At a fixed rms charge radius, the EFG is much less sensitive to the shape of the monopole nuclear charge distribution: electron wavefunctions computed using the Gaussian and Fermi nuclear charge models yield nearly identical EFGs. The recommended effective EFG parameter for the $X\,{}^{2}\Sigma_{1/2}$ electronic ground state is $\langle q\rangle_0=-5.178(21)$~a.u., with an estimated uncertainty reduced by an order of magnitude compared to previous studies. Combined with future measurements of electric-quadrupole hyperfine coupling constants in RaF, this result can be used to determine the nuclear quadrupole moments of Ra isotopes with record precision.
\end{abstract}

\maketitle

\section{Introduction}

Heavy-atom molecules provide enhanced sensitivity to nuclear and particle-physics effects and are therefore useful for precision tests of fundamental symmetries~\cite{KL95,Safronova:18,DeMille2024}. Laser-coolable molecules are particularly promising~\cite{IsaevBerger2016, isaev2017laser,KozyryevDoyle2017,KozyryevHutzler2017,FitchTarbutt2021, Horchani2025}. RaF is attractive because relativistic effects in its electronic structure are large, some Ra nuclei exhibit stable octupole deformation~\cite{Gaffney2013NuclStruc} or octupole vibrations~\cite{Butler:2020}, and its favorable electronic structure supports a practical optical-cycling scheme~\cite{Isaev:2010,Kudashov:14,Udrescu2024}. These properties make RaF relevant both for symmetry-violation searches and for studies of nuclear electromagnetic moments~\cite{wilkins2025observation}.

The first laser spectroscopy of RaF was reported in 2020~\cite{Ruiz:2019}. Subsequent studies measured isotope shifts~\cite{Udrescu2021}, resolved the rotational structure and established a laser-cooling scheme~\cite{Udrescu2024}, and determined the radiative lifetime of the $A\,{}^{2}\Pi_{1/2}$ state~\cite{Athanasakis2024Lifetime}. The excitation energies of fourteen electronic states were measured, and good agreement between relativistic coupled cluster calculations and experiment was found~\cite{Athanasakis2025}. Precision spectroscopy of $^{225}$RaF resolved the Ra hyperfine structure and, together with theoretical input~\cite{Skripnikov:2020e}, enabled the first extraction of the nuclear magnetization distribution effect in a molecule~\cite{wilkins2025observation}. The subpercent agreement between measured and calculated hyperfine parameters provided a stringent probe of the electronic wavefunction near the Ra nucleus.

The $^{225}$Ra nucleus has nuclear spin $I=1/2$ and therefore no electric quadrupole moment. In contrast, $^{223}$Ra has $I=3/2$, so its RaF spectrum contains a contribution from the electric-quadrupole hyperfine coupling in addition to magnetic-dipole terms~\cite{Petrov:2020}. This coupling is conventionally described in terms of the nuclear quadrupole moment and the electric field gradient (EFG) at the Ra nucleus that probes the near-nuclear electron wavefunction, complementing magnetic hyperfine measurements.
The relativistic electronic wavefunction depends on the radial shape of the finite nuclear charge distribution, and so does the EFG at the center of the nucleus. A separate finite-size correction arises when the electronic density penetrates the finite nuclear quadrupole distribution. The latter correction is analogous to the Bohr--Weisskopf effect, the correction to magnetic-dipole hyperfine splitting due to the finite nuclear magnetization distribution~\cite{BohrWeisskopf:1950}. Nuclear size effects on quadrupole coupling in atoms, molecules, and solids were considered in Refs.~\cite{Pyykko1970NuclearVolume, Koch2010Penetration, Filatov2012QuadrupoleAnomaly, RinkerSteffen1977,Steffen1981Muon, Steffen1985Muon,Michel2017Muonic,DzubaFlambaum2023FNS,Jankovic2026Hyperfine} (see also references therein). The relative importance of this correction among the contributions and uncertainties in high-precision molecular EFG calculations has received only modest attention. RaF is a valuable system for such studies: experimental measurements are in progress~\cite{wilkins2025observation}, and it contains one of the heaviest elements, Ra ($Z=88$), with a large nucleus, so both relativistic and nuclear-size effects are enhanced.

The precision with which the nuclear quadrupole moments of Ra isotopes can be obtained from atomic hyperfine measurements is currently limited by uncertainties in the calculated EFGs~\cite{Li2021RaPlus,BieronPyykko2005Ra,Nezosi2026RaHFS}. In principle, these moments can also be extracted from molecular hyperfine measurements combined with calculated EFGs. The electric-quadrupole coupling constant of $^{223}$RaF was previously calculated with an uncertainty of 5\%~\cite{Petrov:2020}. Here, we aim to reduce this uncertainty by an order of magnitude, providing the electronic-structure input needed for precise determinations of nuclear electric quadrupole moments in Ra isotopes. In addition, we investigate the effect of electron penetration into the nucleus and compare its magnitude with those of other contributions that are rarely included in molecular EFG calculations.

\section{Theory}

The Coulomb interaction of the electrons with the nucleus is given by
\begin{equation}
  H_{\rm eN}
 =\iint\frac{ \rho_N(\bm R) \rho_e(\bm r)}
 {|\bm r-\bm R|}\,\mathrm d^3R\,\mathrm d^3r,
 \label{eq:coulomb_gaussian}
\end{equation}
where $ \rho_N$ and $ \rho_e$ are the nuclear and electronic charge-density operators, respectively. Here $\bm r$ is an electronic coordinate and $\bm R$ is a nuclear charge coordinate, both measured from the Ra center. The electric quadrupole (E2) part of this interaction is obtained from the corresponding term in the standard expansion:
\begin{equation}
 \frac{1}{|\bm r-\bm R|}
 =\sum_{\ell m}C_m^{(\ell)}(\widehat{\bm R})
 C_m^{(\ell)*}(\widehat{\bm r})\frac{r_<^{\ell}}{r_>^{\ell+1}},
 \label{eq:coulomb_multipoles}
\end{equation}
where $C_m^{(\ell)}=\sqrt{\frac{4\pi}{2\ell+1}}Y_{\ell m}$, $r_<$ and $r_>$ are the smaller and larger of the radial coordinates $r$ and $R$, and hats on coordinate vectors denote unit directions.

Within a fixed nuclear spin multiplet with $I\geq1$, electronic averaging in an axially symmetric system gives, to first order in the E2 interaction~\cite{BrownCarrington2003},
\begin{equation}
  H_Q^{X}(R_{\mathrm{RaF}})
 =\frac{eQq_X(R_{\mathrm{RaF}})}{4I(2I-1)}
 \bigl[3 I_z^{\,2}-I(I+1)\bigr].
 \label{eq:quadrupole_hamiltonian}
\end{equation}
Here $I_z$ is the nuclear spin projection on the molecular axis. The spectroscopic quadrupole moment $Q$ is defined in the stretched nuclear state by
\begin{equation}
 eQ=2\left\langle I,I\left|
 \int \rho_N(\bm R)R^2C_0^{(2)}(\widehat{\bm R})\,\mathrm d^3R
 \right|I,I\right\rangle,
 \label{eq:Q_spectroscopic}
\end{equation}
where $e>0$ is the elementary charge and $I$ is the Ra nuclear spin.

The molecular $z$ axis is directed from Ra to F, and the origin is at the Ra nucleus. At an internuclear distance $R_{\mathrm{RaF}}$, the first-order quadrupole coupling in the electronic state $\Psi_\Omega$ is characterized by the molecular quadrupole coupling parameter
\begin{equation}
 q_X(R_{\mathrm{RaF}})=\frac{2Z_{\mathrm F}e}{R_{\mathrm{RaF}}^3}
 -2e\left\langle\Psi_\Omega\left|
 \sum_i\frac{C_0^{(2)}(\widehat{\bm r}_i)}{r_i^3}F_X(r_i)
 \right|\Psi_\Omega\right\rangle,
 \label{eq:vzz_total}
\end{equation}
where $Z_{\mathrm F}=9$ and the first term is the contribution of the F nucleus. The radial function $F_X$ is determined by the nuclear quadrupole-distribution model $X$. The point-quadrupole approximation $F_{\mathrm{point}}(r)=1$ gives the usual electric field gradient at the center of the nucleus. In the thin-shell model, the quadrupole component of the nuclear charge density is confined to the surface of a sphere of radius $R_{\mathrm N}$~\cite{Michel2017Muonic}, giving
\begin{equation}
 F_{\mathrm{shell}}(r)=
 \begin{cases}
 (r/R_{\mathrm N})^5, & r\leq R_{\mathrm N},\\
 1, & r>R_{\mathrm N}.
 \end{cases}
 \label{eq:finite_e2_shell}
\end{equation}
This function accounts for electron penetration in the electric quadrupole interaction. The same radial function is used in Refs.~\cite{DzubaFlambaum2023FNS,Tang2026MgSrBa,Michel2017Muonic}.

For the vibrational ground state of the molecule, the conventional coupling constant in units of frequency is
\begin{equation}
 \frac{eQ\langle q_X\rangle_0}{h}
 =234.9648\,\mathrm{MHz}\,
 \frac{Q}{\mathrm b}\frac{\langle q_X\rangle_0}{\mathrm{a.u.}},
 \label{eq:quadrupole_coupling}
\end{equation}
where $\langle q_X\rangle_0$ denotes the vibrational average, $h$ is Planck's constant, and $\mathrm b$ denotes the barn. All $q_X$ values are reported in atomic units of $e/a_0^3$, with $a_0$ the Bohr radius.

We define the finite-quadrupole-distribution correction as
\begin{equation}
\Delta q_{\rm FQ}=q_{\rm shell}-q_{\rm point},
\label{qPen}
\end{equation}
where $q_{\rm shell}$ is
evaluated using the radial function in Eq.~\eqref{eq:finite_e2_shell}. Both values are evaluated with the same electronic state. Here ``point'' refers only to the quadrupole operator, not to the monopole nuclear charge distribution used to obtain the wavefunction.

To solve the electronic-structure problem, we used the four-component relativistic Hamiltonian
\begin{equation}
 \begin{aligned}
  H&=\Lambda_+\left[
 \sum_i h_D(i)+\sum_{i<j} V_{ee}(i,j)
 \right]\Lambda_+ +V_{\rm NN},\\
  h_D(i)&=c\bm\alpha_i\cdot\bm p_i+\beta_i c^2
 +V_{\rm eN}(\bm r_i),
 \end{aligned}
 \label{eq:rel_hamiltonian}
\end{equation}
where $\Lambda_+$ projects the electronic wavefunction onto the positive-energy states obtained in the Dirac--Hartree--Fock procedure, $c$ is the speed of light, $\bm\alpha_i$ and $\beta_i$ are Dirac matrices, $V_{\rm NN}$ is the repulsion of classical nuclei, and $V_{\rm eN}$ is the electron--nucleus interaction operator generated by the spherical monopole charge distributions of both Ra and F nuclei. For the Dirac--Coulomb (DC) Hamiltonian, the electron--electron interaction is
\begin{equation}
  V_{ee}^{\rm C}(i,j)=\frac{1}{r_{ij}},
 \label{eq:vee_dc}
\end{equation}
where $\bm r_{ij}=\bm r_i-\bm r_j$ and $r_{ij}=|\bm r_{ij}|$. The Dirac--Coulomb--Breit (DCB) approximation corresponds to the use of the full Breit interaction in the zero-frequency limit. The corresponding two-electron operator is
\begin{equation}
 \begin{aligned}
  V_{ee}^{\rm CB}(i,j)
 ={}&\frac{1}{r_{ij}}
 -\frac{(c\bm\alpha_i)\cdot(c\bm\alpha_j)}
 {2c^2r_{ij}}\\
 &-\frac{[(c\bm\alpha_i)\cdot\bm r_{ij}]
 [(c\bm\alpha_j)\cdot\bm r_{ij}]}
 {2c^2r_{ij}^3}.
 \end{aligned}
 \label{eq:vee_dcb}
\end{equation}
The Gaunt term is the magnetic current--current interaction,
\begin{equation}
  V_{\rm G}(i,j)
 =-\frac{(c\bm\alpha_i)\cdot(c\bm\alpha_j)}
 {c^2r_{ij}}
 \label{eq:gaunt}
\end{equation}
usually provides the main part of the Breit interaction.

To describe the interaction of the electrons with a finite-size nucleus, one represents the spherical nuclear charge density, in units of $e$, either by the Gaussian nuclear model~\cite{VisscherDyall1997} or by the two-parameter Fermi distribution
\begin{equation}
 \begin{gathered}
 \rho_{\rm F}(R)=\frac{\rho_0}
 {1+\exp[(R-c_{\rm F})/a]},\\
 4\pi\int_0^\infty R^2\rho_{\rm F}(R)\,\mathrm dR=Z.
 \end{gathered}
 \label{eq:fermi_distribution}
\end{equation}
Here $R$ is measured from the nuclear center, $\rho_0$ ensures normalization to $Z$, and the half-density radius $c_{\rm F}$ is chosen to reproduce the rms charge radius. The diffuseness parameter is $a=t/(4\ln 3)$, with the surface thickness conventionally fixed at $t=2.30$~fm for all nuclei.

\section{Computational details}

The baseline EFG calculation used relativistic coupled cluster theory with single, double, and perturbative triple excitations, CCSD(T)~\cite{Visscher:96a}, and the LBasRaF basis set from Ref.~\cite{Skripnikov:2020e}. This basis combines an extended uncontracted Dyall's AE3Z~\cite{Dyall:2009} basis on Ra with the uncontracted Dyall's AE3Z basis on F; the Ra basis contains $(38s,33p,24d,14f,7g,3h,2i)$ functions. All 97 electrons were correlated, with a virtual-spinor energy cutoff of $10\,000\ E_{\rm h}$. This cutoff allows a balanced treatment of spin polarization and core correlation in Ra~\cite{Skripnikov:17a}. The internuclear distance was fixed at $R_e=4.23$ bohr, as in Ref.~\cite{Skripnikov:2020e}. The Dirac--Coulomb--Gaunt (DCG) Hamiltonian was used in the correlation calculation, with a four-index transformation of the Gaunt interaction in Eq.~(\ref{eq:gaunt}) from the atomic-orbital to the molecular-bispinor basis implemented in this work. The correlation contribution of the non-truncated virtual orbitals above the $10\,000\ E_{\rm h}$ cutoff was estimated at the CCSD(T) level using the DC Hamiltonian and Dyall's AE3Z basis set~\cite{Dyall:2009} for both elements.

The basis-set correction was obtained as the difference between the EFGs calculated with the LBasExtRaF~\cite{Skripnikov:2020e} and LBasRaF basis sets at the CCSD(T) level using the DC Hamiltonian. These calculations correlated 69 electrons, with the Ra $1s$--$3d$ shells frozen, and used a virtual-spinor cutoff of $300\ E_{\rm h}$. LBasExtRaF combines an extended AE4Z~\cite{Dyall:2009} basis on Ra with an AAE4Z~\cite{Dyall:2016} basis on F. In total, the Ra basis contains $(42s,38p,27d,17f,11g,3h,2i)$ uncontracted functions. As the basis-set correction was found to be negligibly small (less than $10^{-3}$~a.u.), we performed an additional calculation with an intermediate-quality basis set, using the Ra basis from LBasExtRaF and the F basis from LBasRaF. The EFG obtained with this basis set was used to estimate the uncertainty due to basis-set variation.

Higher-order correlation corrections were evaluated with the SBas basis from Ref.~\cite{Skripnikov:2020e}, which combines a reoptimized uncontracted Dyall VDZ basis~\cite{Dyall:2009} on Ra with aug-cc-pVDZ-DK on F~\cite{Dunning:89}. The iterative-triples ``CCSDT $-$ CCSD(T)'' correction was obtained as the difference between CCSDT~\cite{Kallay:1} and CCSD(T) results, and the perturbative-quadruples correction as the difference between the results of coupled cluster theory with single, double, triple, and perturbative quadruple excitations, CCSDT(Q)~\cite{Kallay:6}, and the CCSDT results. These calculations correlated the 27 outer-core and valence electrons with a virtual-spinor cutoff of $50\ E_{\rm h}$. We set the uncertainty of the CCSDT $-$ CCSD(T) correction to 100\% of its magnitude. We also used the magnitude of the full $(Q)$ contribution as an estimate of the uncertainty from correlation effects beyond CCSDT(Q).

The finite-field method was used to evaluate EFG values in all cases. The nuclear rms  charge radius of Ra was set to 5.6623~fm following Ref.~\cite{VisscherDyall1997}. Increasing this radius to 5.6973~fm, the recommended value for $^{223}$Ra in Ref.~\cite{ANGELI201369}, changed the EFG by less than 0.001~a.u.

For comparison with experiment, we averaged the calculated $q(R)$ curve over the vibrational ground state,
\begin{equation}
 \begin{aligned}
 \left\langle q\right\rangle_{0}
 & =\int\!\left|\varphi_0(R)\right|^2q(R)\,dR,\\
 \Delta_{\rm vib}
 & =\left\langle q\right\rangle_{0}-q(R_e).
 \end{aligned}
 \label{eq:vib}
\end{equation}
Here $q(R)$ is the EFG at the Ra nucleus as a function of the internuclear distance, and the ground-state vibrational wavefunction $\varphi_0(R)$ was obtained from the potential-energy curve calculated in Ref.~\cite{Athanasakis2025}. For this correction, $q(R)$ was calculated at the CCSD(T) level using the DC Hamiltonian within the non-penetrating approximation $F_{\mathrm{point}}$. The uncertainty in $\Delta_{\rm vib}$ was estimated as the absolute difference between the vibrational corrections calculated at the CCSD(T) and CCSD levels.

The contribution from the residual Breit interaction in the zero-frequency limit in Eq.~(\ref{eq:vee_dcb}) was evaluated at the Dirac--Hartree--Fock (DHF) level as the difference between the EFG values calculated with DHF wavefunctions obtained using the DCB and DCG Hamiltonians, with the Gaunt or full Breit interaction included in the self-consistent procedure. We conservatively (see below) assigned an uncertainty to the ``Breit $-$ Gaunt'' correction equal to its magnitude to account for omitted electron--electron QED effects.

We also estimated the contribution of quantum electrodynamics (QED) effects due to vacuum polarization (VP) and self-energy (SE). The VP correction was described by the Uehling potential, while SE was described by a model operator~\cite{Shabaev:13} in the variant of Ref.~\cite{Skripnikov:2021a}. For this purpose, we compared EFG values calculated at the CCSD(T) level with and without the VP and SE interactions. The contribution was found to be 0.002~a.u. This treatment does not include all QED contributions. Therefore, we omitted this correction from the final EFG value but included its magnitude in the uncertainty budget.

To probe the dependence of the quadrupole coupling constant on the nuclear charge distribution model, we implemented {\it molecular} electron--nucleus attraction integrals for the Fermi distribution~\cite{VisscherDyall1997}. In atomic units, the contribution of the Ra nucleus to $V_{\rm eN}$ in Eq.~\eqref{eq:rel_hamiltonian} was approximated as
\begin{equation}
 \begin{aligned}
 V_{\rm eN}^{\rm F}(\bm r)
 &=-\int\frac{\rho_{\rm F}(r')}
 {|\bm r-\bm r'|}\,d^3r'\\
 &\simeq-\sum_{i=1}^{N_r}\sum_{j=1}^{N_\Omega}
 \frac{w_{ij}^{\rm nuc}}{|\bm r-\bm x_{ij}|},
 \qquad \sum_{ij}w_{ij}^{\rm nuc}=Z,
 \end{aligned}
 \label{eq:fermi_quadrature}
\end{equation}
where the superscript $\mathrm F$ denotes the Fermi model and both $\bm r$ and the nuclear quadrature nodes $\bm x_{ij}=r_i\bm n_j$ are measured from the nuclear center. Here $r_i$ are radial quadrature nodes and $\bm n_j$ are unit vectors specifying the Lebedev directions. The effective charges $w_{ij}^{\rm nuc}$ include the nuclear density, the radial volume factor, and quadrature weights, with an overall normalization to $Z$. The corresponding matrix elements were evaluated as weighted sums of standard point-charge attraction integrals implemented in the {\sc dirac} code~\cite{DIRAC17}. For Ra, the nuclear integration used a direct-product grid of $N_r=256$ Gauss--Legendre radial nodes on $[0,c_F+48a]$ and $N_\Omega=302$ Lebedev angular nodes~\cite{Lebedev1977,LebedevLaikov1999}. Reducing $N_r$ to 128 and $N_\Omega$ to 146 in separate calculations changed the EFG by less than $10^{-6}$~a.u. in each case.

Finally, we applied the electron-penetration correction given by Eq.~(\ref{qPen}), calculated at the CCSD(T) level. In the thin-shell model, we set the shell radius $R_{\mathrm N}$ equal to the rms radius of the monopole nuclear charge distribution.

The final EFG uncertainty was calculated as the square root of the sum of the squared uncertainties of the individual contributions.

Relativistic molecular CCSD(T) calculations were performed with the \textsc{DIRAC} program~\cite{Saue:2020,DIRAC17}. The CCSDT and CCSDT(Q) calculations were carried out with the \textsc{MRCC} program \cite{Kallay:1,MRCC2020}. The residual Breit interaction contribution to the EFG was evaluated using the {\sc pyscf} code~\cite{Sun2020PySCF}. Vibrational averaging was performed with the \textsc{vibrot} module of \textsc{openmolcas}~\cite{LiManni2023OpenMolcas}. Code for the four-index transformation of the Gaunt electron--electron integrals and for evaluating electron--nucleus attraction integrals for the Fermi nuclear charge distribution as well as integrals of the finite-size correction to the electric quadrupole coupling parameter was developed in the present work, extending the {\sc oneprop} code~\cite{Skripnikov:16b}.

\section{Results and discussion}

Table~\ref{Tcontributions} summarizes the contributions to the final vibrationally averaged effective EFG parameter $\langle q\rangle_0$. We first add the fixed-geometry corrections listed in the Table to the baseline CCSD(T) value of $q_{\rm point}(R_e)$ at $R_e=4.23$~bohr, yielding $q_{\rm shell}(R_e)$. The vibrational-averaging correction $\Delta_{\rm vib}$ given by Eq.~\eqref{eq:vib} is then added to obtain $\langle q_{\rm shell}\rangle_0$. Below, we use $q$ as shorthand for $q(R_e)$ when the dependence on the internuclear distance is left implicit. Electron correlation at the CCSD(T) level changes $q$ substantially, by about 17\% relative to the DHF result. The contribution of perturbative quadruple excitations is, however, approximately 20 times smaller than the perturbative-triples correction, demonstrating good convergence with respect to the electron-correlation treatment.

The Gaunt contributions $\Delta q_{\rm Gaunt}$, calculated as differences between the EFG values obtained with the DCG and DC Hamiltonians at the DHF, CCSD, and CCSD(T) levels, are given in Table~\ref{tab:finite_quadrupole_correlation}. The dependence on the correlation treatment suggests an interference between this relativistic correction to the electron--electron interaction and electron correlation. However, this dependence is relatively modest, amounting to about 8\% of the final Gaunt contribution. This supports evaluating the correction due to the residual Breit electron--electron interaction beyond the Gaunt approximation at the DHF level. The Gaunt contribution of 1.2\% for RaF can be compared with the contribution of 1.7\% to the EFG in SrO~\cite{Fabbro2026SrNQM}.

\begin{table}[tbp]
\caption{Contributions to the effective EFG parameter $\left\langle q\right\rangle_{0}$ in the $X\,{}^{2}\Sigma_{1/2}$ ground state of RaF. Values are in atomic units. The finite-quadrupole-distribution correction $\Delta q_{\rm FQ}$ uses the thin-shell operator of Eq.~(\ref{eq:finite_e2_shell}). The DHF, CCSD, and CCSD(T) values already include the Gaunt interaction and the F-nucleus contribution. The DHF and CCSD values are given to illustrate the influence of correlation effects.}
\label{Tcontributions}
\begin{ruledtabular}
\begin{tabular}{lc}
Contribution & Value \\
\hline
DHF & $-$4.433 \\
CCSD & $-$5.366 \\
\\
CCSD(T) & $-$5.195 \\
Basis-set correction & $<0.001(10)$ \\
High virt. & $-$0.002(2) \\
CCSDT $-$ CCSD(T) & 0.009(9) \\
CCSDT(Q) $-$ CCSDT & $-$0.009(9) \\
Breit $-$ Gaunt & $-$0.012(12) \\
Fermi charge distribution &  0.002(1)\\ 
Electr. penetration, $\Delta q_{\rm FQ}$ & 0.009(1) \\
Vibrational averaging, $\Delta_{\rm vib}$ & 0.018(1) \\
Total $\left\langle q\right\rangle_{0}$ & $-5.178(21)$ \\
\\
Ref.~\cite{Petrov:2020} &  $-5.183(259)$\\
\end{tabular}
\end{ruledtabular}
\end{table}

\begin{table}[tbp]
\caption{Dependence of the Gaunt contribution $\Delta q_{\rm Gaunt}$ and the finite-quadrupole-distribution correction $\Delta q_{\rm FQ}$ on the electron-correlation method. Values are in atomic units.}
\label{tab:finite_quadrupole_correlation}
\begin{ruledtabular}
\begin{tabular}{lcc}
Method & $\Delta q_{\rm Gaunt}$ & $\Delta q_{\rm FQ}$ \\
\hline
DHF & 0.067 & 0.009 \\
CCSD & 0.064 & 0.009 \\
CCSD(T) & 0.062 & 0.009 \\
\end{tabular}
\end{ruledtabular}
\end{table}

The dependence of $\Delta q_{\rm FQ}$ on the treatment of electron correlation is shown in Table~\ref{tab:finite_quadrupole_correlation}. All three methods yield $0.009$~a.u. at the reported precision, indicating weak sensitivity to electron correlation, in contrast to the leading point-quadrupole contribution. We conservatively assigned an uncertainty of $0.001$~a.u. to this correction. Uncertainty in the radial distribution of the nuclear quadrupole charge is not included in the quoted uncertainty of the effective $q$. Although $\Delta q_{\rm FQ}=0.009$~a.u. is small relative to the total EFG, its magnitude is comparable to that of the perturbative-quadruples correction, and it must therefore be included at the achieved level of theory. The penetration correction is also comparable in magnitude to the residual Breit correction and is half the vibrational correction. 

The finite-quadrupole-distribution correction can also be compared with the atomic case. Direct diagonal corrections in hydrogen-like systems are small, but in many-electron atoms they can be enhanced by core polarization and correlation through the off-diagonal $s_{1/2}$--$d_{3/2,5/2}$ and $p_{1/2}$--$p_{3/2}$ channels~\cite{DzubaFlambaum2023FNS}. The RaF result is consistent with the relevant partial-wave mixing being present in the molecular DHF spinors because of the anisotropic molecular field.

Another finite-size effect arises from changes in the electronic wavefunction caused by the chosen nuclear charge distribution, specifically through the monopole part of the electron--nucleus interaction in Eq.~(\ref{eq:coulomb_gaussian}). The absolute finite-size effect, calculated as the difference between EFG values obtained with the Gaussian nuclear charge distribution and a point nucleus, is 0.034~a.u. at the DHF level. This effect is about half the DHF Gaunt contribution but nearly four times as large as the electron-penetration correction. However, the radial nuclear-shape effect, calculated as the difference between EFG values obtained with wavefunctions for the Fermi and Gaussian nuclear charge distributions at a fixed rms charge radius within the CCSD(T) model, is only 0.002~a.u., a factor of about 4.5 smaller than the electron-penetration effect.

The final effective molecular EFG, $\langle q \rangle_0=-5.178(21)$~a.u., has a relative theoretical uncertainty of about 0.4\%, an order of magnitude smaller than in the previous molecular study~\cite{Petrov:2020}. This uncertainty can also be compared with the uncertainty estimates reported for atomic EFGs. The relativistic coupled cluster results of Li \textit{et al.} for Ra$^+$ have relative theoretical EFG uncertainties of 1.6--2.2\% for different electronic states~\cite{Li2021RaPlus}. Interestingly, the molecular EFG is about 1.8--3.9 times larger in magnitude than the atomic ones, depending on the chosen atomic state. For neutral Ra, the approximately 2.5\% relative uncertainties in the $^{223}$Ra quadrupole moments extracted using the multiconfiguration Dirac--Hartree--Fock (MCDHF) EFGs of Biero{\'n} and Pyykk{\"o}~\cite{BieronPyykko2005Ra} are dominated by theory. Additional atomic studies of electric-quadrupole hyperfine interactions have been reported~\cite{Heully1985RaHFS,Andriessen1992RaHFS,Bieron2005RaHFS,Sahoo2007RaPlus,HassounehSalah2022RaHFS}, but separate uncertainty estimates for the electronic factors are not specified. Recent MCDHF calculations by Nezosi \textit{et al.} report $B$ values with estimated $2\sigma$ uncertainties exceeding 5\% for different electronic states of Ra$^+$ and Ra~\cite{Nezosi2026RaHFS}. At the same time, the relative experimental uncertainties in the electric-quadrupole hyperfine constants $B$ for selected states of Ra and Ra$^+$ are approximately one to two orders of magnitude smaller~\cite{Wendt1987RaHFS,Neu1989RaIIQuadrupole,Lynch2018RaCRIS}. Thus, uncertainties in atomic theory currently limit the precision with which the spectroscopic quadrupole moments of Ra isotopes can be extracted from atomic hyperfine measurements. A precise measurement of the electric-quadrupole coupling constant in $^{223}$RaF, combined with the present effective molecular EFG, would enable a more precise determination of the spectroscopic quadrupole moment of $^{223}$Ra.

\section{Conclusion}

We determined the vibrationally averaged effective EFG parameter at the Ra nucleus in ground-state RaF using four-component relativistic coupled cluster theory. The recommended value, $\langle q \rangle_0=-5.178(21)$~a.u., includes the separately evaluated correction for electron penetration into the finite nuclear quadrupole distribution, which is non-negligible at the achieved precision. Within the adopted thin-shell nuclear quadrupole distribution model, this result provides the electronic input for extracting Ra nuclear quadrupole moments from measured electric-quadrupole hyperfine couplings in the RaF molecule. In particular, a measurement in $^{223}$RaF could yield a reference value of $Q(^{223}\mathrm{Ra})$ that, together with measured quadrupole hyperfine constants $B$ in $^{223}$Ra and $^{223}$Ra$^+$, would enable an semiempirical calibration of the state-specific atomic $B/Q$ factors. These calibrated factors could then be used to determine quadrupole moments of other Ra isotopes from their measured $B$ constants in the same electronic states.
Our comparison of higher-order electron correlation, Breit, QED, nuclear-shape, and finite-quadrupole-distribution effects quantifies their relative contributions to the effective electric-field gradient in RaF and can guide future studies of other systems.

\section{Acknowledgments}
Electronic structure calculations were carried out using computing resources of the federal collective usage center Complex for Simulation and Data Processing for Mega-science Facilities at National Research Centre ``Kurchatov Institute'', http://ckp.nrcki.ru/, and partly using the computing resources of the quantum chemistry laboratory. The research has been supported by the Russian Science Foundation Grant No. 26-12-00350 \url{https://rscf.ru/project/26-12-00350/}.

\section*{Conflict of interest}
The authors have no conflicts to disclose.

\section*{DATA AVAILABILITY}The data that support the findings of this study are available from the corresponding author upon reasonable request.


\begin{thebibliography}{64}%
\makeatletter
\providecommand \@ifxundefined [1]{%
 \@ifx{#1\undefined}
}%
\providecommand \@ifnum [1]{%
 \ifnum #1\expandafter \@firstoftwo
 \else \expandafter \@secondoftwo
 \fi
}%
\providecommand \@ifx [1]{%
 \ifx #1\expandafter \@firstoftwo
 \else \expandafter \@secondoftwo
 \fi
}%
\providecommand \natexlab [1]{#1}%
\providecommand \enquote  [1]{``#1''}%
\providecommand \bibnamefont  [1]{#1}%
\providecommand \bibfnamefont [1]{#1}%
\providecommand \citenamefont [1]{#1}%
\providecommand \href@noop [0]{\@secondoftwo}%
\providecommand \href [0]{\begingroup \@sanitize@url \@href}%
\providecommand \@href[1]{\@@startlink{#1}\@@href}%
\providecommand \@@href[1]{\endgroup#1\@@endlink}%
\providecommand \@sanitize@url [0]{\catcode `\\12\catcode `\$12\catcode `\&12\catcode `\#12\catcode `\^12\catcode `\_12\catcode `\%12\relax}%
\providecommand \@@startlink[1]{}%
\providecommand \@@endlink[0]{}%
\providecommand \url  [0]{\begingroup\@sanitize@url \@url }%
\providecommand \@url [1]{\endgroup\@href {#1}{\urlprefix }}%
\providecommand \urlprefix  [0]{URL }%
\providecommand \Eprint [0]{\href }%
\providecommand \doibase [0]{https://doi.org/}%
\providecommand \selectlanguage [0]{\@gobble}%
\providecommand \bibinfo  [0]{\@secondoftwo}%
\providecommand \bibfield  [0]{\@secondoftwo}%
\providecommand \translation [1]{[#1]}%
\providecommand \BibitemOpen [0]{}%
\providecommand \bibitemStop [0]{}%
\providecommand \bibitemNoStop [0]{.\EOS\space}%
\providecommand \EOS [0]{\spacefactor3000\relax}%
\providecommand \BibitemShut  [1]{\csname bibitem#1\endcsname}%
\let\auto@bib@innerbib\@empty
\bibitem [{\citenamefont {Kozlov}\ and\ \citenamefont {Labzowsky}(1995)}]{KL95}%
  \BibitemOpen
  \bibfield  {author} {\bibinfo {author} {\bibfnamefont {M.~G.}\ \bibnamefont {Kozlov}}\ and\ \bibinfo {author} {\bibfnamefont {L.~N.}\ \bibnamefont {Labzowsky}},\ }\href {https://doi.org/10.1088/0953-4075/28/10/008} {\bibfield  {journal} {\bibinfo  {journal} {J. Phys. B: At. Mol. Opt. Phys.}\ }\textbf {\bibinfo {volume} {28}},\ \bibinfo {pages} {1933} (\bibinfo {year} {1995})}\BibitemShut {NoStop}%
\bibitem [{\citenamefont {Safronova}\ \emph {et~al.}(2018)\citenamefont {Safronova}, \citenamefont {Budker}, \citenamefont {DeMille}, \citenamefont {Jackson~Kimball}, \citenamefont {Derevianko},\ and\ \citenamefont {Clark}}]{Safronova:18}%
  \BibitemOpen
  \bibfield  {author} {\bibinfo {author} {\bibfnamefont {M.~S.}\ \bibnamefont {Safronova}}, \bibinfo {author} {\bibfnamefont {D.}~\bibnamefont {Budker}}, \bibinfo {author} {\bibfnamefont {D.}~\bibnamefont {DeMille}}, \bibinfo {author} {\bibfnamefont {D.~F.}\ \bibnamefont {Jackson~Kimball}}, \bibinfo {author} {\bibfnamefont {A.}~\bibnamefont {Derevianko}},\ and\ \bibinfo {author} {\bibfnamefont {C.~W.}\ \bibnamefont {Clark}},\ }\href {https://doi.org/10.1103/RevModPhys.90.025008} {\bibfield  {journal} {\bibinfo  {journal} {Rev.\ Mod.\ Phys.}\ }\textbf {\bibinfo {volume} {90}},\ \bibinfo {pages} {025008} (\bibinfo {year} {2018})}\BibitemShut {NoStop}%
\bibitem [{\citenamefont {DeMille}\ \emph {et~al.}(2024)\citenamefont {DeMille}, \citenamefont {Hutzler}, \citenamefont {Rey},\ and\ \citenamefont {Zelevinsky}}]{DeMille2024}%
  \BibitemOpen
  \bibfield  {author} {\bibinfo {author} {\bibfnamefont {D.}~\bibnamefont {DeMille}}, \bibinfo {author} {\bibfnamefont {N.~R.}\ \bibnamefont {Hutzler}}, \bibinfo {author} {\bibfnamefont {A.~M.}\ \bibnamefont {Rey}},\ and\ \bibinfo {author} {\bibfnamefont {T.}~\bibnamefont {Zelevinsky}},\ }\href {https://doi.org/10.1038/s41567-024-02499-9} {\bibfield  {journal} {\bibinfo  {journal} {Nat. Phys.}\ }\textbf {\bibinfo {volume} {20}},\ \bibinfo {pages} {741} (\bibinfo {year} {2024})}\BibitemShut {NoStop}%
\bibitem [{\citenamefont {Isaev}\ and\ \citenamefont {Berger}(2016)}]{IsaevBerger2016}%
  \BibitemOpen
  \bibfield  {author} {\bibinfo {author} {\bibfnamefont {T.~A.}\ \bibnamefont {Isaev}}\ and\ \bibinfo {author} {\bibfnamefont {R.}~\bibnamefont {Berger}},\ }\href {https://doi.org/10.1103/PhysRevLett.116.063006} {\bibfield  {journal} {\bibinfo  {journal} {Phys. Rev. Lett.}\ }\textbf {\bibinfo {volume} {116}},\ \bibinfo {pages} {063006} (\bibinfo {year} {2016})}\BibitemShut {NoStop}%
\bibitem [{\citenamefont {Isaev}\ \emph {et~al.}(2017)\citenamefont {Isaev}, \citenamefont {Zaitsevskii},\ and\ \citenamefont {Eliav}}]{isaev2017laser}%
  \BibitemOpen
  \bibfield  {author} {\bibinfo {author} {\bibfnamefont {T.~A.}\ \bibnamefont {Isaev}}, \bibinfo {author} {\bibfnamefont {A.~V.}\ \bibnamefont {Zaitsevskii}},\ and\ \bibinfo {author} {\bibfnamefont {E.}~\bibnamefont {Eliav}},\ }\href {https://doi.org/10.1088/1361-6455/aa8f34} {\bibfield  {journal} {\bibinfo  {journal} {J. Phys. B: At. Mol. Opt. Phys.}\ }\textbf {\bibinfo {volume} {50}},\ \bibinfo {pages} {225101} (\bibinfo {year} {2017})}\BibitemShut {NoStop}%
\bibitem [{\citenamefont {Kozyryev}\ \emph {et~al.}(2017)\citenamefont {Kozyryev}, \citenamefont {Baum}, \citenamefont {Matsuda}, \citenamefont {Augenbraun}, \citenamefont {Anderegg}, \citenamefont {Sedlack},\ and\ \citenamefont {Doyle}}]{KozyryevDoyle2017}%
  \BibitemOpen
  \bibfield  {author} {\bibinfo {author} {\bibfnamefont {I.}~\bibnamefont {Kozyryev}}, \bibinfo {author} {\bibfnamefont {L.}~\bibnamefont {Baum}}, \bibinfo {author} {\bibfnamefont {K.}~\bibnamefont {Matsuda}}, \bibinfo {author} {\bibfnamefont {B.~L.}\ \bibnamefont {Augenbraun}}, \bibinfo {author} {\bibfnamefont {L.}~\bibnamefont {Anderegg}}, \bibinfo {author} {\bibfnamefont {A.~P.}\ \bibnamefont {Sedlack}},\ and\ \bibinfo {author} {\bibfnamefont {J.~M.}\ \bibnamefont {Doyle}},\ }\href {https://doi.org/10.1103/PhysRevLett.118.173201} {\bibfield  {journal} {\bibinfo  {journal} {Phys. Rev. Lett.}\ }\textbf {\bibinfo {volume} {118}},\ \bibinfo {pages} {173201} (\bibinfo {year} {2017})}\BibitemShut {NoStop}%
\bibitem [{\citenamefont {Kozyryev}\ and\ \citenamefont {Hutzler}(2017)}]{KozyryevHutzler2017}%
  \BibitemOpen
  \bibfield  {author} {\bibinfo {author} {\bibfnamefont {I.}~\bibnamefont {Kozyryev}}\ and\ \bibinfo {author} {\bibfnamefont {N.~R.}\ \bibnamefont {Hutzler}},\ }\href {https://doi.org/10.1103/PhysRevLett.119.133002} {\bibfield  {journal} {\bibinfo  {journal} {Phys. Rev. Lett.}\ }\textbf {\bibinfo {volume} {119}},\ \bibinfo {pages} {133002} (\bibinfo {year} {2017})}\BibitemShut {NoStop}%
\bibitem [{\citenamefont {Fitch}\ and\ \citenamefont {Tarbutt}(2021)}]{FitchTarbutt2021}%
  \BibitemOpen
  \bibfield  {author} {\bibinfo {author} {\bibfnamefont {N.~J.}\ \bibnamefont {Fitch}}\ and\ \bibinfo {author} {\bibfnamefont {M.~R.}\ \bibnamefont {Tarbutt}},\ }in\ \href {https://doi.org/10.1016/bs.aamop.2021.04.003} {\emph {\bibinfo {booktitle} {Adv. At. Mol. Opt. Phys.}}},\ Vol.~\bibinfo {volume} {70}\ (\bibinfo  {publisher} {Academic Press},\ \bibinfo {year} {2021})\ pp.\ \bibinfo {pages} {157--262}\BibitemShut {NoStop}%
\bibitem [{\citenamefont {Horchani}\ and\ \citenamefont {Al-Shaqsi}(2025)}]{Horchani2025}%
  \BibitemOpen
  \bibfield  {author} {\bibinfo {author} {\bibfnamefont {R.}~\bibnamefont {Horchani}}\ and\ \bibinfo {author} {\bibfnamefont {A.}~\bibnamefont {Al-Shaqsi}},\ }\href {https://doi.org/10.1116/5.0235255} {\bibfield  {journal} {\bibinfo  {journal} {AVS Quantum Sci.}\ }\textbf {\bibinfo {volume} {7}},\ \bibinfo {pages} {013501} (\bibinfo {year} {2025})}\BibitemShut {NoStop}%
\bibitem [{\citenamefont {Gaffney}\ \emph {et~al.}(2013)\citenamefont {Gaffney}, \citenamefont {Butler}, \citenamefont {Scheck}, \citenamefont {Hayes}, \citenamefont {Wenander}, \citenamefont {Albers}, \citenamefont {Bastin}, \citenamefont {Bauer}, \citenamefont {Blazhev}, \citenamefont {B{\"{o}}nig}, \citenamefont {Bree}, \citenamefont {Cederk{\"{a}}ll}, \citenamefont {Chupp}, \citenamefont {Cline}, \citenamefont {Cocolios}, \citenamefont {Davinson}, \citenamefont {De~Witte}, \citenamefont {Diriken}, \citenamefont {Grahn}, \citenamefont {Herzan}, \citenamefont {Huyse}, \citenamefont {Jenkins}, \citenamefont {Joss}, \citenamefont {Kesteloot}, \citenamefont {Konki}, \citenamefont {Kowalczyk}, \citenamefont {Kr{\"{o}}ll}, \citenamefont {Kwan}, \citenamefont {Lutter}, \citenamefont {Moschner}, \citenamefont {Napiorkowski}, \citenamefont {Pakarinen}, \citenamefont {Pfeiffer}, \citenamefont {Radeck}, \citenamefont {Reiter}, \citenamefont {Reynders}, \citenamefont {Rigby}, \citenamefont {Robledo}, \citenamefont
  {Rudigier}, \citenamefont {Sambi}, \citenamefont {Seidlitz}, \citenamefont {Siebeck}, \citenamefont {Stora}, \citenamefont {Thoele}, \citenamefont {Van~Duppen}, \citenamefont {Vermeulen}, \citenamefont {Von~Schmid}, \citenamefont {Voulot}, \citenamefont {Warr}, \citenamefont {Wimmer}, \citenamefont {Wrzosek-Lipska}, \citenamefont {Wu},\ and\ \citenamefont {Zielinska}}]{Gaffney2013NuclStruc}%
  \BibitemOpen
  \bibfield  {author} {\bibinfo {author} {\bibfnamefont {L.~P.}\ \bibnamefont {Gaffney}}, \bibinfo {author} {\bibfnamefont {P.~A.}\ \bibnamefont {Butler}}, \bibinfo {author} {\bibfnamefont {M.}~\bibnamefont {Scheck}}, \bibinfo {author} {\bibfnamefont {A.~B.}\ \bibnamefont {Hayes}}, \bibinfo {author} {\bibfnamefont {F.}~\bibnamefont {Wenander}}, \bibinfo {author} {\bibfnamefont {M.}~\bibnamefont {Albers}}, \bibinfo {author} {\bibfnamefont {B.}~\bibnamefont {Bastin}}, \bibinfo {author} {\bibfnamefont {C.}~\bibnamefont {Bauer}}, \bibinfo {author} {\bibfnamefont {A.}~\bibnamefont {Blazhev}}, \bibinfo {author} {\bibfnamefont {S.}~\bibnamefont {B{\"{o}}nig}}, \bibinfo {author} {\bibfnamefont {N.}~\bibnamefont {Bree}}, \bibinfo {author} {\bibfnamefont {J.}~\bibnamefont {Cederk{\"{a}}ll}}, \bibinfo {author} {\bibfnamefont {T.}~\bibnamefont {Chupp}}, \bibinfo {author} {\bibfnamefont {D.}~\bibnamefont {Cline}}, \bibinfo {author} {\bibfnamefont {T.~E.}\ \bibnamefont {Cocolios}}, \bibinfo {author} {\bibfnamefont
  {T.}~\bibnamefont {Davinson}}, \bibinfo {author} {\bibfnamefont {H.}~\bibnamefont {De~Witte}}, \bibinfo {author} {\bibfnamefont {J.}~\bibnamefont {Diriken}}, \bibinfo {author} {\bibfnamefont {T.}~\bibnamefont {Grahn}}, \bibinfo {author} {\bibfnamefont {A.}~\bibnamefont {Herzan}}, \bibinfo {author} {\bibfnamefont {M.}~\bibnamefont {Huyse}}, \bibinfo {author} {\bibfnamefont {D.~G.}\ \bibnamefont {Jenkins}}, \bibinfo {author} {\bibfnamefont {D.~T.}\ \bibnamefont {Joss}}, \bibinfo {author} {\bibfnamefont {N.}~\bibnamefont {Kesteloot}}, \bibinfo {author} {\bibfnamefont {J.}~\bibnamefont {Konki}}, \bibinfo {author} {\bibfnamefont {M.}~\bibnamefont {Kowalczyk}}, \bibinfo {author} {\bibfnamefont {T.}~\bibnamefont {Kr{\"{o}}ll}}, \bibinfo {author} {\bibfnamefont {E.}~\bibnamefont {Kwan}}, \bibinfo {author} {\bibfnamefont {R.}~\bibnamefont {Lutter}}, \bibinfo {author} {\bibfnamefont {K.}~\bibnamefont {Moschner}}, \bibinfo {author} {\bibfnamefont {P.}~\bibnamefont {Napiorkowski}}, \bibinfo {author} {\bibfnamefont
  {J.}~\bibnamefont {Pakarinen}}, \bibinfo {author} {\bibfnamefont {M.}~\bibnamefont {Pfeiffer}}, \bibinfo {author} {\bibfnamefont {D.}~\bibnamefont {Radeck}}, \bibinfo {author} {\bibfnamefont {P.}~\bibnamefont {Reiter}}, \bibinfo {author} {\bibfnamefont {K.}~\bibnamefont {Reynders}}, \bibinfo {author} {\bibfnamefont {S.~V.}\ \bibnamefont {Rigby}}, \bibinfo {author} {\bibfnamefont {L.~M.}\ \bibnamefont {Robledo}}, \bibinfo {author} {\bibfnamefont {M.}~\bibnamefont {Rudigier}}, \bibinfo {author} {\bibfnamefont {S.}~\bibnamefont {Sambi}}, \bibinfo {author} {\bibfnamefont {M.}~\bibnamefont {Seidlitz}}, \bibinfo {author} {\bibfnamefont {B.}~\bibnamefont {Siebeck}}, \bibinfo {author} {\bibfnamefont {T.}~\bibnamefont {Stora}}, \bibinfo {author} {\bibfnamefont {P.}~\bibnamefont {Thoele}}, \bibinfo {author} {\bibfnamefont {P.}~\bibnamefont {Van~Duppen}}, \bibinfo {author} {\bibfnamefont {M.~J.}\ \bibnamefont {Vermeulen}}, \bibinfo {author} {\bibfnamefont {M.}~\bibnamefont {Von~Schmid}}, \bibinfo {author}
  {\bibfnamefont {D.}~\bibnamefont {Voulot}}, \bibinfo {author} {\bibfnamefont {N.}~\bibnamefont {Warr}}, \bibinfo {author} {\bibfnamefont {K.}~\bibnamefont {Wimmer}}, \bibinfo {author} {\bibfnamefont {K.}~\bibnamefont {Wrzosek-Lipska}}, \bibinfo {author} {\bibfnamefont {C.~Y.}\ \bibnamefont {Wu}},\ and\ \bibinfo {author} {\bibfnamefont {M.}~\bibnamefont {Zielinska}},\ }\href {https://doi.org/10.1038/nature12073} {\bibfield  {journal} {\bibinfo  {journal} {Nature}\ }\textbf {\bibinfo {volume} {497}},\ \bibinfo {pages} {199} (\bibinfo {year} {2013})}\BibitemShut {NoStop}%
\bibitem [{\citenamefont {Butler}\ \emph {et~al.}(2020)\citenamefont {Butler}, \citenamefont {Gaffney}, \citenamefont {Spagnoletti}, \citenamefont {Abrahams}, \citenamefont {Bowry}, \citenamefont {Cederk\"all}, \citenamefont {de~Angelis}, \citenamefont {De~Witte}, \citenamefont {Garrett}, \citenamefont {Goldkuhle}, \citenamefont {Henrich}, \citenamefont {Illana}, \citenamefont {Johnston}, \citenamefont {Joss}, \citenamefont {Keatings}, \citenamefont {Kelly}, \citenamefont {Komorowska}, \citenamefont {Konki}, \citenamefont {Kr\"oll}, \citenamefont {Lozano}, \citenamefont {Nara~Singh}, \citenamefont {O'Donnell}, \citenamefont {Ojala}, \citenamefont {Page}, \citenamefont {Pedersen}, \citenamefont {Raison}, \citenamefont {Reiter}, \citenamefont {Rodriguez}, \citenamefont {Rosiak}, \citenamefont {Rothe}, \citenamefont {Scheck}, \citenamefont {Seidlitz}, \citenamefont {Shneidman}, \citenamefont {Siebeck}, \citenamefont {Sinclair}, \citenamefont {Smith}, \citenamefont {Stryjczyk}, \citenamefont {Van~Duppen},
  \citenamefont {Vinals}, \citenamefont {Virtanen}, \citenamefont {Warr}, \citenamefont {Wrzosek-Lipska},\ and\ \citenamefont {Zieli\ifmmode~\acute{n}\else \'{n}\fi{}ska}}]{Butler:2020}%
  \BibitemOpen
  \bibfield  {author} {\bibinfo {author} {\bibfnamefont {P.~A.}\ \bibnamefont {Butler}}, \bibinfo {author} {\bibfnamefont {L.~P.}\ \bibnamefont {Gaffney}}, \bibinfo {author} {\bibfnamefont {P.}~\bibnamefont {Spagnoletti}}, \bibinfo {author} {\bibfnamefont {K.}~\bibnamefont {Abrahams}}, \bibinfo {author} {\bibfnamefont {M.}~\bibnamefont {Bowry}}, \bibinfo {author} {\bibfnamefont {J.}~\bibnamefont {Cederk\"all}}, \bibinfo {author} {\bibfnamefont {G.}~\bibnamefont {de~Angelis}}, \bibinfo {author} {\bibfnamefont {H.}~\bibnamefont {De~Witte}}, \bibinfo {author} {\bibfnamefont {P.~E.}\ \bibnamefont {Garrett}}, \bibinfo {author} {\bibfnamefont {A.}~\bibnamefont {Goldkuhle}}, \bibinfo {author} {\bibfnamefont {C.}~\bibnamefont {Henrich}}, \bibinfo {author} {\bibfnamefont {A.}~\bibnamefont {Illana}}, \bibinfo {author} {\bibfnamefont {K.}~\bibnamefont {Johnston}}, \bibinfo {author} {\bibfnamefont {D.~T.}\ \bibnamefont {Joss}}, \bibinfo {author} {\bibfnamefont {J.~M.}\ \bibnamefont {Keatings}}, \bibinfo {author}
  {\bibfnamefont {N.~A.}\ \bibnamefont {Kelly}}, \bibinfo {author} {\bibfnamefont {M.}~\bibnamefont {Komorowska}}, \bibinfo {author} {\bibfnamefont {J.}~\bibnamefont {Konki}}, \bibinfo {author} {\bibfnamefont {T.}~\bibnamefont {Kr\"oll}}, \bibinfo {author} {\bibfnamefont {M.}~\bibnamefont {Lozano}}, \bibinfo {author} {\bibfnamefont {B.~S.}\ \bibnamefont {Nara~Singh}}, \bibinfo {author} {\bibfnamefont {D.}~\bibnamefont {O'Donnell}}, \bibinfo {author} {\bibfnamefont {J.}~\bibnamefont {Ojala}}, \bibinfo {author} {\bibfnamefont {R.~D.}\ \bibnamefont {Page}}, \bibinfo {author} {\bibfnamefont {L.~G.}\ \bibnamefont {Pedersen}}, \bibinfo {author} {\bibfnamefont {C.}~\bibnamefont {Raison}}, \bibinfo {author} {\bibfnamefont {P.}~\bibnamefont {Reiter}}, \bibinfo {author} {\bibfnamefont {J.~A.}\ \bibnamefont {Rodriguez}}, \bibinfo {author} {\bibfnamefont {D.}~\bibnamefont {Rosiak}}, \bibinfo {author} {\bibfnamefont {S.}~\bibnamefont {Rothe}}, \bibinfo {author} {\bibfnamefont {M.}~\bibnamefont {Scheck}}, \bibinfo {author}
  {\bibfnamefont {M.}~\bibnamefont {Seidlitz}}, \bibinfo {author} {\bibfnamefont {T.~M.}\ \bibnamefont {Shneidman}}, \bibinfo {author} {\bibfnamefont {B.}~\bibnamefont {Siebeck}}, \bibinfo {author} {\bibfnamefont {J.}~\bibnamefont {Sinclair}}, \bibinfo {author} {\bibfnamefont {J.~F.}\ \bibnamefont {Smith}}, \bibinfo {author} {\bibfnamefont {M.}~\bibnamefont {Stryjczyk}}, \bibinfo {author} {\bibfnamefont {P.}~\bibnamefont {Van~Duppen}}, \bibinfo {author} {\bibfnamefont {S.}~\bibnamefont {Vinals}}, \bibinfo {author} {\bibfnamefont {V.}~\bibnamefont {Virtanen}}, \bibinfo {author} {\bibfnamefont {N.}~\bibnamefont {Warr}}, \bibinfo {author} {\bibfnamefont {K.}~\bibnamefont {Wrzosek-Lipska}},\ and\ \bibinfo {author} {\bibfnamefont {M.}~\bibnamefont {Zieli\ifmmode~\acute{n}\else \'{n}\fi{}ska}},\ }\href {https://doi.org/10.1103/PhysRevLett.124.042503} {\bibfield  {journal} {\bibinfo  {journal} {Phys.\ Rev.\ Lett.}\ }\textbf {\bibinfo {volume} {124}},\ \bibinfo {pages} {042503} (\bibinfo {year} {2020})}\BibitemShut
  {NoStop}%
\bibitem [{\citenamefont {Isaev}\ \emph {et~al.}(2010)\citenamefont {Isaev}, \citenamefont {Hoekstra},\ and\ \citenamefont {Berger}}]{Isaev:2010}%
  \BibitemOpen
  \bibfield  {author} {\bibinfo {author} {\bibfnamefont {T.~A.}\ \bibnamefont {Isaev}}, \bibinfo {author} {\bibfnamefont {S.}~\bibnamefont {Hoekstra}},\ and\ \bibinfo {author} {\bibfnamefont {R.}~\bibnamefont {Berger}},\ }\href {https://doi.org/10.1103/PhysRevA.82.052521} {\bibfield  {journal} {\bibinfo  {journal} {Phys.\ Rev.\ A}\ }\textbf {\bibinfo {volume} {82}},\ \bibinfo {pages} {052521} (\bibinfo {year} {2010})}\BibitemShut {NoStop}%
\bibitem [{\citenamefont {Kudashov}\ \emph {et~al.}(2014)\citenamefont {Kudashov}, \citenamefont {Petrov}, \citenamefont {Skripnikov}, \citenamefont {Mosyagin}, \citenamefont {Isaev}, \citenamefont {Berger},\ and\ \citenamefont {Titov}}]{Kudashov:14}%
  \BibitemOpen
  \bibfield  {author} {\bibinfo {author} {\bibfnamefont {A.~D.}\ \bibnamefont {Kudashov}}, \bibinfo {author} {\bibfnamefont {A.~N.}\ \bibnamefont {Petrov}}, \bibinfo {author} {\bibfnamefont {L.~V.}\ \bibnamefont {Skripnikov}}, \bibinfo {author} {\bibfnamefont {N.~S.}\ \bibnamefont {Mosyagin}}, \bibinfo {author} {\bibfnamefont {T.~A.}\ \bibnamefont {Isaev}}, \bibinfo {author} {\bibfnamefont {R.}~\bibnamefont {Berger}},\ and\ \bibinfo {author} {\bibfnamefont {A.~V.}\ \bibnamefont {Titov}},\ }\href {https://doi.org/10.1103/PhysRevA.90.052513} {\bibfield  {journal} {\bibinfo  {journal} {Phys. Rev. A}\ }\textbf {\bibinfo {volume} {90}},\ \bibinfo {pages} {052513} (\bibinfo {year} {2014})}\BibitemShut {NoStop}%
\bibitem [{\citenamefont {Udrescu}\ \emph {et~al.}(2024)\citenamefont {Udrescu}, \citenamefont {Wilkins}, \citenamefont {Breier}, \citenamefont {Athanasakis-Kaklamanakis}, \citenamefont {Garcia~Ruiz}, \citenamefont {Au}, \citenamefont {Belo{\v{s}}evi{\'c}}, \citenamefont {Berger}, \citenamefont {Bissell}, \citenamefont {Binnersley}, \citenamefont {Brinson}, \citenamefont {Chrysalidis}, \citenamefont {Cocolios}, \citenamefont {de~Groote}, \citenamefont {Dorne}, \citenamefont {Flanagan}, \citenamefont {Franchoo}, \citenamefont {Gaul}, \citenamefont {Geldhof}, \citenamefont {Giesen}, \citenamefont {Hanstorp}, \citenamefont {Heinke}, \citenamefont {Koszor{\'u}s}, \citenamefont {Kujanp{\"a}{\"a}}, \citenamefont {Lalanne}, \citenamefont {Neyens}, \citenamefont {Nichols}, \citenamefont {Perrett}, \citenamefont {Reilly}, \citenamefont {Rothe}, \citenamefont {van~den Borne}, \citenamefont {Vernon}, \citenamefont {Wang}, \citenamefont {Wessolek}, \citenamefont {Yang},\ and\ \citenamefont {Z{\"u}lch}}]{Udrescu2024}%
  \BibitemOpen
  \bibfield  {author} {\bibinfo {author} {\bibfnamefont {S.~M.}\ \bibnamefont {Udrescu}}, \bibinfo {author} {\bibfnamefont {S.~G.}\ \bibnamefont {Wilkins}}, \bibinfo {author} {\bibfnamefont {A.~A.}\ \bibnamefont {Breier}}, \bibinfo {author} {\bibfnamefont {M.}~\bibnamefont {Athanasakis-Kaklamanakis}}, \bibinfo {author} {\bibfnamefont {R.~F.}\ \bibnamefont {Garcia~Ruiz}}, \bibinfo {author} {\bibfnamefont {M.}~\bibnamefont {Au}}, \bibinfo {author} {\bibfnamefont {I.}~\bibnamefont {Belo{\v{s}}evi{\'c}}}, \bibinfo {author} {\bibfnamefont {R.}~\bibnamefont {Berger}}, \bibinfo {author} {\bibfnamefont {M.~L.}\ \bibnamefont {Bissell}}, \bibinfo {author} {\bibfnamefont {C.~L.}\ \bibnamefont {Binnersley}}, \bibinfo {author} {\bibfnamefont {A.~J.}\ \bibnamefont {Brinson}}, \bibinfo {author} {\bibfnamefont {K.}~\bibnamefont {Chrysalidis}}, \bibinfo {author} {\bibfnamefont {T.~E.}\ \bibnamefont {Cocolios}}, \bibinfo {author} {\bibfnamefont {R.~P.}\ \bibnamefont {de~Groote}}, \bibinfo {author} {\bibfnamefont
  {A.}~\bibnamefont {Dorne}}, \bibinfo {author} {\bibfnamefont {K.~T.}\ \bibnamefont {Flanagan}}, \bibinfo {author} {\bibfnamefont {S.}~\bibnamefont {Franchoo}}, \bibinfo {author} {\bibfnamefont {K.}~\bibnamefont {Gaul}}, \bibinfo {author} {\bibfnamefont {S.}~\bibnamefont {Geldhof}}, \bibinfo {author} {\bibfnamefont {T.~F.}\ \bibnamefont {Giesen}}, \bibinfo {author} {\bibfnamefont {D.}~\bibnamefont {Hanstorp}}, \bibinfo {author} {\bibfnamefont {R.}~\bibnamefont {Heinke}}, \bibinfo {author} {\bibfnamefont {{\'A}.}~\bibnamefont {Koszor{\'u}s}}, \bibinfo {author} {\bibfnamefont {S.}~\bibnamefont {Kujanp{\"a}{\"a}}}, \bibinfo {author} {\bibfnamefont {L.}~\bibnamefont {Lalanne}}, \bibinfo {author} {\bibfnamefont {G.}~\bibnamefont {Neyens}}, \bibinfo {author} {\bibfnamefont {M.}~\bibnamefont {Nichols}}, \bibinfo {author} {\bibfnamefont {H.~A.}\ \bibnamefont {Perrett}}, \bibinfo {author} {\bibfnamefont {J.~R.}\ \bibnamefont {Reilly}}, \bibinfo {author} {\bibfnamefont {S.}~\bibnamefont {Rothe}}, \bibinfo {author}
  {\bibfnamefont {B.}~\bibnamefont {van~den Borne}}, \bibinfo {author} {\bibfnamefont {A.~R.}\ \bibnamefont {Vernon}}, \bibinfo {author} {\bibfnamefont {Q.}~\bibnamefont {Wang}}, \bibinfo {author} {\bibfnamefont {J.}~\bibnamefont {Wessolek}}, \bibinfo {author} {\bibfnamefont {X.~F.}\ \bibnamefont {Yang}},\ and\ \bibinfo {author} {\bibfnamefont {C.}~\bibnamefont {Z{\"u}lch}},\ }\href {https://doi.org/10.1038/s41567-023-02296-w} {\bibfield  {journal} {\bibinfo  {journal} {Nat. Phys.}\ }\textbf {\bibinfo {volume} {20}},\ \bibinfo {pages} {202} (\bibinfo {year} {2024})}\BibitemShut {NoStop}%
\bibitem [{\citenamefont {Wilkins}\ \emph {et~al.}(2025)\citenamefont {Wilkins}, \citenamefont {Udrescu}, \citenamefont {Athanasakis-Kaklamanakis}, \citenamefont {Ruiz}, \citenamefont {Au}, \citenamefont {Belo\v{s}evi\'{c}}, \citenamefont {Berger}, \citenamefont {Bissell}, \citenamefont {Breier}, \citenamefont {Brinson}, \citenamefont {Chrysalidis}, \citenamefont {Cocolios}, \citenamefont {de~Groote}, \citenamefont {Dorne}, \citenamefont {Flanagan}, \citenamefont {Franchoo}, \citenamefont {Gaul}, \citenamefont {Geldhof}, \citenamefont {Giesen}, \citenamefont {Hanstorp}, \citenamefont {Heinke}, \citenamefont {Isaev}, \citenamefont {Koszor\'{u}s}, \citenamefont {Kujanp\"{a}\"{a}}, \citenamefont {Lalanne}, \citenamefont {Neyens}, \citenamefont {Nichols}, \citenamefont {Perrett}, \citenamefont {Reilly}, \citenamefont {Skripnikov}, \citenamefont {Rothe}, \citenamefont {van~den Borne}, \citenamefont {Wang}, \citenamefont {Wessolek}, \citenamefont {Yang},\ and\ \citenamefont {Z\"{u}lch}}]{wilkins2025observation}%
  \BibitemOpen
  \bibfield  {author} {\bibinfo {author} {\bibfnamefont {S.~G.}\ \bibnamefont {Wilkins}}, \bibinfo {author} {\bibfnamefont {S.~M.}\ \bibnamefont {Udrescu}}, \bibinfo {author} {\bibfnamefont {M.}~\bibnamefont {Athanasakis-Kaklamanakis}}, \bibinfo {author} {\bibfnamefont {R.~F.~G.}\ \bibnamefont {Ruiz}}, \bibinfo {author} {\bibfnamefont {M.}~\bibnamefont {Au}}, \bibinfo {author} {\bibfnamefont {I.}~\bibnamefont {Belo\v{s}evi\'{c}}}, \bibinfo {author} {\bibfnamefont {R.}~\bibnamefont {Berger}}, \bibinfo {author} {\bibfnamefont {M.~L.}\ \bibnamefont {Bissell}}, \bibinfo {author} {\bibfnamefont {A.~A.}\ \bibnamefont {Breier}}, \bibinfo {author} {\bibfnamefont {A.~J.}\ \bibnamefont {Brinson}}, \bibinfo {author} {\bibfnamefont {K.}~\bibnamefont {Chrysalidis}}, \bibinfo {author} {\bibfnamefont {T.~E.}\ \bibnamefont {Cocolios}}, \bibinfo {author} {\bibfnamefont {R.~P.}\ \bibnamefont {de~Groote}}, \bibinfo {author} {\bibfnamefont {A.}~\bibnamefont {Dorne}}, \bibinfo {author} {\bibfnamefont {K.~T.}\ \bibnamefont
  {Flanagan}}, \bibinfo {author} {\bibfnamefont {S.}~\bibnamefont {Franchoo}}, \bibinfo {author} {\bibfnamefont {K.}~\bibnamefont {Gaul}}, \bibinfo {author} {\bibfnamefont {S.}~\bibnamefont {Geldhof}}, \bibinfo {author} {\bibfnamefont {T.~F.}\ \bibnamefont {Giesen}}, \bibinfo {author} {\bibfnamefont {D.}~\bibnamefont {Hanstorp}}, \bibinfo {author} {\bibfnamefont {R.}~\bibnamefont {Heinke}}, \bibinfo {author} {\bibfnamefont {T.}~\bibnamefont {Isaev}}, \bibinfo {author} {\bibfnamefont {A.}~\bibnamefont {Koszor\'{u}s}}, \bibinfo {author} {\bibfnamefont {S.}~\bibnamefont {Kujanp\"{a}\"{a}}}, \bibinfo {author} {\bibfnamefont {L.}~\bibnamefont {Lalanne}}, \bibinfo {author} {\bibfnamefont {G.}~\bibnamefont {Neyens}}, \bibinfo {author} {\bibfnamefont {M.}~\bibnamefont {Nichols}}, \bibinfo {author} {\bibfnamefont {H.~A.}\ \bibnamefont {Perrett}}, \bibinfo {author} {\bibfnamefont {J.~R.}\ \bibnamefont {Reilly}}, \bibinfo {author} {\bibfnamefont {L.~V.}\ \bibnamefont {Skripnikov}}, \bibinfo {author} {\bibfnamefont
  {S.}~\bibnamefont {Rothe}}, \bibinfo {author} {\bibfnamefont {B.}~\bibnamefont {van~den Borne}}, \bibinfo {author} {\bibfnamefont {Q.}~\bibnamefont {Wang}}, \bibinfo {author} {\bibfnamefont {J.}~\bibnamefont {Wessolek}}, \bibinfo {author} {\bibfnamefont {X.~F.}\ \bibnamefont {Yang}},\ and\ \bibinfo {author} {\bibfnamefont {C.}~\bibnamefont {Z\"{u}lch}},\ }\href {https://doi.org/10.1126/science.adm7717} {\bibfield  {journal} {\bibinfo  {journal} {Science}\ }\textbf {\bibinfo {volume} {390}},\ \bibinfo {pages} {386} (\bibinfo {year} {2025})}\BibitemShut {NoStop}%
\bibitem [{\citenamefont {Garcia~Ruiz}\ \emph {et~al.}(2020)\citenamefont {Garcia~Ruiz}, \citenamefont {Berger}, \citenamefont {Billowes}, \citenamefont {Binnersley}, \citenamefont {Bissell}, \citenamefont {Breier}, \citenamefont {Brinson}, \citenamefont {Chrysalidis}, \citenamefont {Cocolios}, \citenamefont {Cooper}, \citenamefont {Flanagan}, \citenamefont {Giesen}, \citenamefont {de~Groote}, \citenamefont {Franchoo}, \citenamefont {Gustafsson}, \citenamefont {Isaev}, \citenamefont {Koszor{\'u}s}, \citenamefont {Neyens}, \citenamefont {Perrett}, \citenamefont {Ricketts}, \citenamefont {Rothe}, \citenamefont {Schweikhard}, \citenamefont {Vernon}, \citenamefont {Wendt}, \citenamefont {Wienholtz}, \citenamefont {Wilkins},\ and\ \citenamefont {Yang}}]{Ruiz:2019}%
  \BibitemOpen
  \bibfield  {author} {\bibinfo {author} {\bibfnamefont {R.~F.}\ \bibnamefont {Garcia~Ruiz}}, \bibinfo {author} {\bibfnamefont {R.}~\bibnamefont {Berger}}, \bibinfo {author} {\bibfnamefont {J.}~\bibnamefont {Billowes}}, \bibinfo {author} {\bibfnamefont {C.~L.}\ \bibnamefont {Binnersley}}, \bibinfo {author} {\bibfnamefont {M.~L.}\ \bibnamefont {Bissell}}, \bibinfo {author} {\bibfnamefont {A.~A.}\ \bibnamefont {Breier}}, \bibinfo {author} {\bibfnamefont {A.~J.}\ \bibnamefont {Brinson}}, \bibinfo {author} {\bibfnamefont {K.}~\bibnamefont {Chrysalidis}}, \bibinfo {author} {\bibfnamefont {T.~E.}\ \bibnamefont {Cocolios}}, \bibinfo {author} {\bibfnamefont {B.~S.}\ \bibnamefont {Cooper}}, \bibinfo {author} {\bibfnamefont {K.~T.}\ \bibnamefont {Flanagan}}, \bibinfo {author} {\bibfnamefont {T.~F.}\ \bibnamefont {Giesen}}, \bibinfo {author} {\bibfnamefont {R.~P.}\ \bibnamefont {de~Groote}}, \bibinfo {author} {\bibfnamefont {S.}~\bibnamefont {Franchoo}}, \bibinfo {author} {\bibfnamefont {F.~P.}\ \bibnamefont
  {Gustafsson}}, \bibinfo {author} {\bibfnamefont {T.~A.}\ \bibnamefont {Isaev}}, \bibinfo {author} {\bibfnamefont {{\'A}.}~\bibnamefont {Koszor{\'u}s}}, \bibinfo {author} {\bibfnamefont {G.}~\bibnamefont {Neyens}}, \bibinfo {author} {\bibfnamefont {H.~A.}\ \bibnamefont {Perrett}}, \bibinfo {author} {\bibfnamefont {C.~M.}\ \bibnamefont {Ricketts}}, \bibinfo {author} {\bibfnamefont {S.}~\bibnamefont {Rothe}}, \bibinfo {author} {\bibfnamefont {L.}~\bibnamefont {Schweikhard}}, \bibinfo {author} {\bibfnamefont {A.~R.}\ \bibnamefont {Vernon}}, \bibinfo {author} {\bibfnamefont {K.~D.~A.}\ \bibnamefont {Wendt}}, \bibinfo {author} {\bibfnamefont {F.}~\bibnamefont {Wienholtz}}, \bibinfo {author} {\bibfnamefont {S.~G.}\ \bibnamefont {Wilkins}},\ and\ \bibinfo {author} {\bibfnamefont {X.~F.}\ \bibnamefont {Yang}},\ }\href {https://doi.org/10.1038/s41586-020-2299-4} {\bibfield  {journal} {\bibinfo  {journal} {Nature}\ }\textbf {\bibinfo {volume} {581}},\ \bibinfo {pages} {396} (\bibinfo {year} {2020})}\BibitemShut
  {NoStop}%
\bibitem [{\citenamefont {Udrescu}\ \emph {et~al.}(2021)\citenamefont {Udrescu}, \citenamefont {Brinson}, \citenamefont {Garcia~Ruiz}, \citenamefont {Gaul}, \citenamefont {Berger}, \citenamefont {Billowes}, \citenamefont {Binnersley}, \citenamefont {Bissell}, \citenamefont {Breier}, \citenamefont {Chrysalidis}, \citenamefont {Cocolios}, \citenamefont {Cooper}, \citenamefont {Flanagan}, \citenamefont {Giesen}, \citenamefont {de~Groote}, \citenamefont {Franchoo}, \citenamefont {Gustafsson}, \citenamefont {Isaev}, \citenamefont {Koszor{\'u}s}, \citenamefont {Neyens}, \citenamefont {Perrett}, \citenamefont {Ricketts}, \citenamefont {Rothe}, \citenamefont {Vernon}, \citenamefont {Wendt}, \citenamefont {Wienholtz}, \citenamefont {Wilkins},\ and\ \citenamefont {Yang}}]{Udrescu2021}%
  \BibitemOpen
  \bibfield  {author} {\bibinfo {author} {\bibfnamefont {S.~M.}\ \bibnamefont {Udrescu}}, \bibinfo {author} {\bibfnamefont {A.~J.}\ \bibnamefont {Brinson}}, \bibinfo {author} {\bibfnamefont {R.~F.}\ \bibnamefont {Garcia~Ruiz}}, \bibinfo {author} {\bibfnamefont {K.}~\bibnamefont {Gaul}}, \bibinfo {author} {\bibfnamefont {R.}~\bibnamefont {Berger}}, \bibinfo {author} {\bibfnamefont {J.}~\bibnamefont {Billowes}}, \bibinfo {author} {\bibfnamefont {C.~L.}\ \bibnamefont {Binnersley}}, \bibinfo {author} {\bibfnamefont {M.~L.}\ \bibnamefont {Bissell}}, \bibinfo {author} {\bibfnamefont {A.~A.}\ \bibnamefont {Breier}}, \bibinfo {author} {\bibfnamefont {K.}~\bibnamefont {Chrysalidis}}, \bibinfo {author} {\bibfnamefont {T.~E.}\ \bibnamefont {Cocolios}}, \bibinfo {author} {\bibfnamefont {B.~S.}\ \bibnamefont {Cooper}}, \bibinfo {author} {\bibfnamefont {K.~T.}\ \bibnamefont {Flanagan}}, \bibinfo {author} {\bibfnamefont {T.~F.}\ \bibnamefont {Giesen}}, \bibinfo {author} {\bibfnamefont {R.~P.}\ \bibnamefont {de~Groote}},
  \bibinfo {author} {\bibfnamefont {S.}~\bibnamefont {Franchoo}}, \bibinfo {author} {\bibfnamefont {F.~P.}\ \bibnamefont {Gustafsson}}, \bibinfo {author} {\bibfnamefont {T.~A.}\ \bibnamefont {Isaev}}, \bibinfo {author} {\bibfnamefont {{\'A}.}~\bibnamefont {Koszor{\'u}s}}, \bibinfo {author} {\bibfnamefont {G.}~\bibnamefont {Neyens}}, \bibinfo {author} {\bibfnamefont {H.~A.}\ \bibnamefont {Perrett}}, \bibinfo {author} {\bibfnamefont {C.~M.}\ \bibnamefont {Ricketts}}, \bibinfo {author} {\bibfnamefont {S.}~\bibnamefont {Rothe}}, \bibinfo {author} {\bibfnamefont {A.~R.}\ \bibnamefont {Vernon}}, \bibinfo {author} {\bibfnamefont {K.~D.~A.}\ \bibnamefont {Wendt}}, \bibinfo {author} {\bibfnamefont {F.}~\bibnamefont {Wienholtz}}, \bibinfo {author} {\bibfnamefont {S.~G.}\ \bibnamefont {Wilkins}},\ and\ \bibinfo {author} {\bibfnamefont {X.~F.}\ \bibnamefont {Yang}},\ }\href {https://doi.org/10.1103/PhysRevLett.127.033001} {\bibfield  {journal} {\bibinfo  {journal} {Phys. Rev. Lett.}\ }\textbf {\bibinfo {volume} {127}},\
  \bibinfo {pages} {033001} (\bibinfo {year} {2021})}\BibitemShut {NoStop}%
\bibitem [{\citenamefont {Athanasakis-Kaklamanakis}\ \emph {et~al.}(2024)\citenamefont {Athanasakis-Kaklamanakis}, \citenamefont {Wilkins}, \citenamefont {Lass{\`e}gues}, \citenamefont {Lalanne}, \citenamefont {Reilly}, \citenamefont {Ahmad}, \citenamefont {Au}, \citenamefont {Bai}, \citenamefont {Berbalk}, \citenamefont {Bernerd}, \citenamefont {Borschevsky}, \citenamefont {Breier}, \citenamefont {Chrysalidis}, \citenamefont {Cocolios}, \citenamefont {de~Groote}, \citenamefont {Fajardo-Zambrano}, \citenamefont {Flanagan}, \citenamefont {Franchoo}, \citenamefont {Garcia~Ruiz}, \citenamefont {Hanstorp}, \citenamefont {Heinke}, \citenamefont {Imgram}, \citenamefont {Koszor{\'u}s}, \citenamefont {Kyuberis}, \citenamefont {Lim}, \citenamefont {Liu}, \citenamefont {Lynch}, \citenamefont {McGlone}, \citenamefont {Mei}, \citenamefont {Neyens}, \citenamefont {Nies}, \citenamefont {Oleynichenko}, \citenamefont {Raggio}, \citenamefont {Rothe}, \citenamefont {Skripnikov}, \citenamefont {Smets}, \citenamefont {van~den
  Borne}, \citenamefont {Warbinek}, \citenamefont {Wessolek},\ and\ \citenamefont {Yang}}]{Athanasakis2024Lifetime}%
  \BibitemOpen
  \bibfield  {author} {\bibinfo {author} {\bibfnamefont {M.}~\bibnamefont {Athanasakis-Kaklamanakis}}, \bibinfo {author} {\bibfnamefont {S.~G.}\ \bibnamefont {Wilkins}}, \bibinfo {author} {\bibfnamefont {P.}~\bibnamefont {Lass{\`e}gues}}, \bibinfo {author} {\bibfnamefont {L.}~\bibnamefont {Lalanne}}, \bibinfo {author} {\bibfnamefont {J.~R.}\ \bibnamefont {Reilly}}, \bibinfo {author} {\bibfnamefont {O.}~\bibnamefont {Ahmad}}, \bibinfo {author} {\bibfnamefont {M.}~\bibnamefont {Au}}, \bibinfo {author} {\bibfnamefont {S.~W.}\ \bibnamefont {Bai}}, \bibinfo {author} {\bibfnamefont {J.}~\bibnamefont {Berbalk}}, \bibinfo {author} {\bibfnamefont {C.}~\bibnamefont {Bernerd}}, \bibinfo {author} {\bibfnamefont {A.}~\bibnamefont {Borschevsky}}, \bibinfo {author} {\bibfnamefont {A.~A.}\ \bibnamefont {Breier}}, \bibinfo {author} {\bibfnamefont {K.}~\bibnamefont {Chrysalidis}}, \bibinfo {author} {\bibfnamefont {T.~E.}\ \bibnamefont {Cocolios}}, \bibinfo {author} {\bibfnamefont {R.~P.}\ \bibnamefont {de~Groote}}, \bibinfo
  {author} {\bibfnamefont {C.~M.}\ \bibnamefont {Fajardo-Zambrano}}, \bibinfo {author} {\bibfnamefont {K.~T.}\ \bibnamefont {Flanagan}}, \bibinfo {author} {\bibfnamefont {S.}~\bibnamefont {Franchoo}}, \bibinfo {author} {\bibfnamefont {R.~F.}\ \bibnamefont {Garcia~Ruiz}}, \bibinfo {author} {\bibfnamefont {D.}~\bibnamefont {Hanstorp}}, \bibinfo {author} {\bibfnamefont {R.}~\bibnamefont {Heinke}}, \bibinfo {author} {\bibfnamefont {P.}~\bibnamefont {Imgram}}, \bibinfo {author} {\bibfnamefont {{\'A}.}~\bibnamefont {Koszor{\'u}s}}, \bibinfo {author} {\bibfnamefont {A.~A.}\ \bibnamefont {Kyuberis}}, \bibinfo {author} {\bibfnamefont {J.}~\bibnamefont {Lim}}, \bibinfo {author} {\bibfnamefont {Y.~C.}\ \bibnamefont {Liu}}, \bibinfo {author} {\bibfnamefont {K.~M.}\ \bibnamefont {Lynch}}, \bibinfo {author} {\bibfnamefont {A.}~\bibnamefont {McGlone}}, \bibinfo {author} {\bibfnamefont {W.~C.}\ \bibnamefont {Mei}}, \bibinfo {author} {\bibfnamefont {G.}~\bibnamefont {Neyens}}, \bibinfo {author} {\bibfnamefont
  {L.}~\bibnamefont {Nies}}, \bibinfo {author} {\bibfnamefont {A.~V.}\ \bibnamefont {Oleynichenko}}, \bibinfo {author} {\bibfnamefont {A.}~\bibnamefont {Raggio}}, \bibinfo {author} {\bibfnamefont {S.}~\bibnamefont {Rothe}}, \bibinfo {author} {\bibfnamefont {L.~V.}\ \bibnamefont {Skripnikov}}, \bibinfo {author} {\bibfnamefont {E.}~\bibnamefont {Smets}}, \bibinfo {author} {\bibfnamefont {B.}~\bibnamefont {van~den Borne}}, \bibinfo {author} {\bibfnamefont {J.}~\bibnamefont {Warbinek}}, \bibinfo {author} {\bibfnamefont {J.}~\bibnamefont {Wessolek}},\ and\ \bibinfo {author} {\bibfnamefont {X.~F.}\ \bibnamefont {Yang}},\ }\href {https://doi.org/10.1103/PhysRevA.110.L010802} {\bibfield  {journal} {\bibinfo  {journal} {Phys. Rev. A}\ }\textbf {\bibinfo {volume} {110}},\ \bibinfo {pages} {L010802} (\bibinfo {year} {2024})}\BibitemShut {NoStop}%
\bibitem [{\citenamefont {Athanasakis-Kaklamanakis}\ \emph {et~al.}(2025)\citenamefont {Athanasakis-Kaklamanakis}, \citenamefont {Wilkins}, \citenamefont {Skripnikov}, \citenamefont {Koszor{\'u}s}, \citenamefont {Breier}, \citenamefont {Ahmad}, \citenamefont {Au}, \citenamefont {Bai}, \citenamefont {Belo{\v{s}}evi{\'c}}, \citenamefont {Berbalk}, \citenamefont {Berger}, \citenamefont {Bernerd}, \citenamefont {Bissell}, \citenamefont {Borschevsky}, \citenamefont {Brinson}, \citenamefont {Chrysalidis}, \citenamefont {Cocolios}, \citenamefont {de~Groote}, \citenamefont {Dorne}, \citenamefont {Fajardo-Zambrano}, \citenamefont {Field}, \citenamefont {Flanagan}, \citenamefont {Franchoo}, \citenamefont {Garcia~Ruiz}, \citenamefont {Gaul}, \citenamefont {Geldhof}, \citenamefont {Giesen}, \citenamefont {Hanstorp}, \citenamefont {Heinke}, \citenamefont {Imgram}, \citenamefont {Isaev}, \citenamefont {Kyuberis}, \citenamefont {Kujanp{\"a}{\"a}}, \citenamefont {Lalanne}, \citenamefont {Lass{\`e}gues}, \citenamefont {Lim},
  \citenamefont {Liu}, \citenamefont {Lynch}, \citenamefont {McGlone}, \citenamefont {Mei}, \citenamefont {Neyens}, \citenamefont {Nichols}, \citenamefont {Nies}, \citenamefont {Pa{\v{s}}teka}, \citenamefont {Perrett}, \citenamefont {Raggio}, \citenamefont {Reilly}, \citenamefont {Rothe}, \citenamefont {Smets}, \citenamefont {Udrescu}, \citenamefont {van~den Borne}, \citenamefont {Wang}, \citenamefont {Warbinek}, \citenamefont {Wessolek}, \citenamefont {Yang},\ and\ \citenamefont {Z{\"u}lch}}]{Athanasakis2025}%
  \BibitemOpen
  \bibfield  {author} {\bibinfo {author} {\bibfnamefont {M.}~\bibnamefont {Athanasakis-Kaklamanakis}}, \bibinfo {author} {\bibfnamefont {S.~G.}\ \bibnamefont {Wilkins}}, \bibinfo {author} {\bibfnamefont {L.~V.}\ \bibnamefont {Skripnikov}}, \bibinfo {author} {\bibfnamefont {{\'A}.}~\bibnamefont {Koszor{\'u}s}}, \bibinfo {author} {\bibfnamefont {A.~A.}\ \bibnamefont {Breier}}, \bibinfo {author} {\bibfnamefont {O.}~\bibnamefont {Ahmad}}, \bibinfo {author} {\bibfnamefont {M.}~\bibnamefont {Au}}, \bibinfo {author} {\bibfnamefont {S.~W.}\ \bibnamefont {Bai}}, \bibinfo {author} {\bibfnamefont {I.}~\bibnamefont {Belo{\v{s}}evi{\'c}}}, \bibinfo {author} {\bibfnamefont {J.}~\bibnamefont {Berbalk}}, \bibinfo {author} {\bibfnamefont {R.}~\bibnamefont {Berger}}, \bibinfo {author} {\bibfnamefont {C.}~\bibnamefont {Bernerd}}, \bibinfo {author} {\bibfnamefont {M.~L.}\ \bibnamefont {Bissell}}, \bibinfo {author} {\bibfnamefont {A.}~\bibnamefont {Borschevsky}}, \bibinfo {author} {\bibfnamefont {A.}~\bibnamefont {Brinson}},
  \bibinfo {author} {\bibfnamefont {K.}~\bibnamefont {Chrysalidis}}, \bibinfo {author} {\bibfnamefont {T.~E.}\ \bibnamefont {Cocolios}}, \bibinfo {author} {\bibfnamefont {R.~P.}\ \bibnamefont {de~Groote}}, \bibinfo {author} {\bibfnamefont {A.}~\bibnamefont {Dorne}}, \bibinfo {author} {\bibfnamefont {C.~M.}\ \bibnamefont {Fajardo-Zambrano}}, \bibinfo {author} {\bibfnamefont {R.~W.}\ \bibnamefont {Field}}, \bibinfo {author} {\bibfnamefont {K.~T.}\ \bibnamefont {Flanagan}}, \bibinfo {author} {\bibfnamefont {S.}~\bibnamefont {Franchoo}}, \bibinfo {author} {\bibfnamefont {R.~F.}\ \bibnamefont {Garcia~Ruiz}}, \bibinfo {author} {\bibfnamefont {K.}~\bibnamefont {Gaul}}, \bibinfo {author} {\bibfnamefont {S.}~\bibnamefont {Geldhof}}, \bibinfo {author} {\bibfnamefont {T.~F.}\ \bibnamefont {Giesen}}, \bibinfo {author} {\bibfnamefont {D.}~\bibnamefont {Hanstorp}}, \bibinfo {author} {\bibfnamefont {R.}~\bibnamefont {Heinke}}, \bibinfo {author} {\bibfnamefont {P.}~\bibnamefont {Imgram}}, \bibinfo {author} {\bibfnamefont
  {T.~A.}\ \bibnamefont {Isaev}}, \bibinfo {author} {\bibfnamefont {A.~A.}\ \bibnamefont {Kyuberis}}, \bibinfo {author} {\bibfnamefont {S.}~\bibnamefont {Kujanp{\"a}{\"a}}}, \bibinfo {author} {\bibfnamefont {L.}~\bibnamefont {Lalanne}}, \bibinfo {author} {\bibfnamefont {P.}~\bibnamefont {Lass{\`e}gues}}, \bibinfo {author} {\bibfnamefont {J.}~\bibnamefont {Lim}}, \bibinfo {author} {\bibfnamefont {Y.~C.}\ \bibnamefont {Liu}}, \bibinfo {author} {\bibfnamefont {K.~M.}\ \bibnamefont {Lynch}}, \bibinfo {author} {\bibfnamefont {A.}~\bibnamefont {McGlone}}, \bibinfo {author} {\bibfnamefont {W.~C.}\ \bibnamefont {Mei}}, \bibinfo {author} {\bibfnamefont {G.}~\bibnamefont {Neyens}}, \bibinfo {author} {\bibfnamefont {M.}~\bibnamefont {Nichols}}, \bibinfo {author} {\bibfnamefont {L.}~\bibnamefont {Nies}}, \bibinfo {author} {\bibfnamefont {L.~F.}\ \bibnamefont {Pa{\v{s}}teka}}, \bibinfo {author} {\bibfnamefont {H.~A.}\ \bibnamefont {Perrett}}, \bibinfo {author} {\bibfnamefont {A.}~\bibnamefont {Raggio}}, \bibinfo {author}
  {\bibfnamefont {J.~R.}\ \bibnamefont {Reilly}}, \bibinfo {author} {\bibfnamefont {S.}~\bibnamefont {Rothe}}, \bibinfo {author} {\bibfnamefont {E.}~\bibnamefont {Smets}}, \bibinfo {author} {\bibfnamefont {S.-M.}\ \bibnamefont {Udrescu}}, \bibinfo {author} {\bibfnamefont {B.}~\bibnamefont {van~den Borne}}, \bibinfo {author} {\bibfnamefont {Q.}~\bibnamefont {Wang}}, \bibinfo {author} {\bibfnamefont {J.}~\bibnamefont {Warbinek}}, \bibinfo {author} {\bibfnamefont {J.}~\bibnamefont {Wessolek}}, \bibinfo {author} {\bibfnamefont {X.~F.}\ \bibnamefont {Yang}},\ and\ \bibinfo {author} {\bibfnamefont {C.}~\bibnamefont {Z{\"u}lch}},\ }\href {https://doi.org/10.1038/s41467-025-55977-w} {\bibfield  {journal} {\bibinfo  {journal} {Nat. Commun.}\ }\textbf {\bibinfo {volume} {16}},\ \bibinfo {pages} {2139} (\bibinfo {year} {2025})}\BibitemShut {NoStop}%
\bibitem [{\citenamefont {Skripnikov}(2020)}]{Skripnikov:2020e}%
  \BibitemOpen
  \bibfield  {author} {\bibinfo {author} {\bibfnamefont {L.~V.}\ \bibnamefont {Skripnikov}},\ }\href {https://doi.org/10.1063/5.0024103} {\bibfield  {journal} {\bibinfo  {journal} {J. Chem. Phys.}\ }\textbf {\bibinfo {volume} {153}},\ \bibinfo {pages} {114114} (\bibinfo {year} {2020})}\BibitemShut {NoStop}%
\bibitem [{\citenamefont {Petrov}\ and\ \citenamefont {Skripnikov}(2020)}]{Petrov:2020}%
  \BibitemOpen
  \bibfield  {author} {\bibinfo {author} {\bibfnamefont {A.~N.}\ \bibnamefont {Petrov}}\ and\ \bibinfo {author} {\bibfnamefont {L.~V.}\ \bibnamefont {Skripnikov}},\ }\href {https://doi.org/10.1103/PhysRevA.102.062801} {\bibfield  {journal} {\bibinfo  {journal} {Phys. Rev. A}\ }\textbf {\bibinfo {volume} {102}},\ \bibinfo {pages} {062801} (\bibinfo {year} {2020})}\BibitemShut {NoStop}%
\bibitem [{\citenamefont {Bohr}\ and\ \citenamefont {Weisskopf}(1950)}]{BohrWeisskopf:1950}%
  \BibitemOpen
  \bibfield  {author} {\bibinfo {author} {\bibfnamefont {A.}~\bibnamefont {Bohr}}\ and\ \bibinfo {author} {\bibfnamefont {V.~F.}\ \bibnamefont {Weisskopf}},\ }\href {https://doi.org/10.1103/PhysRev.77.94} {\bibfield  {journal} {\bibinfo  {journal} {Phys. Rev.}\ }\textbf {\bibinfo {volume} {77}},\ \bibinfo {pages} {94} (\bibinfo {year} {1950})}\BibitemShut {NoStop}%
\bibitem [{\citenamefont {Pyykk{\"o}}(1970)}]{Pyykko1970NuclearVolume}%
  \BibitemOpen
  \bibfield  {author} {\bibinfo {author} {\bibfnamefont {P.}~\bibnamefont {Pyykk{\"o}}},\ }\href {https://doi.org/10.1016/0009-2614(70)85197-1} {\bibfield  {journal} {\bibinfo  {journal} {Chem. Phys. Lett.}\ }\textbf {\bibinfo {volume} {6}},\ \bibinfo {pages} {479} (\bibinfo {year} {1970})}\BibitemShut {NoStop}%
\bibitem [{\citenamefont {Koch}\ \emph {et~al.}(2010)\citenamefont {Koch}, \citenamefont {Koepernik}, \citenamefont {Van~Neck}, \citenamefont {Rosner},\ and\ \citenamefont {Cottenier}}]{Koch2010Penetration}%
  \BibitemOpen
  \bibfield  {author} {\bibinfo {author} {\bibfnamefont {K.}~\bibnamefont {Koch}}, \bibinfo {author} {\bibfnamefont {K.}~\bibnamefont {Koepernik}}, \bibinfo {author} {\bibfnamefont {D.}~\bibnamefont {Van~Neck}}, \bibinfo {author} {\bibfnamefont {H.}~\bibnamefont {Rosner}},\ and\ \bibinfo {author} {\bibfnamefont {S.}~\bibnamefont {Cottenier}},\ }\href {https://doi.org/10.1103/PhysRevA.81.032507} {\bibfield  {journal} {\bibinfo  {journal} {Phys. Rev. A}\ }\textbf {\bibinfo {volume} {81}},\ \bibinfo {pages} {032507} (\bibinfo {year} {2010})}\BibitemShut {NoStop}%
\bibitem [{\citenamefont {Filatov}\ \emph {et~al.}(2012)\citenamefont {Filatov}, \citenamefont {Zou},\ and\ \citenamefont {Cremer}}]{Filatov2012QuadrupoleAnomaly}%
  \BibitemOpen
  \bibfield  {author} {\bibinfo {author} {\bibfnamefont {M.}~\bibnamefont {Filatov}}, \bibinfo {author} {\bibfnamefont {W.}~\bibnamefont {Zou}},\ and\ \bibinfo {author} {\bibfnamefont {D.}~\bibnamefont {Cremer}},\ }\href {https://doi.org/10.1063/1.4757568} {\bibfield  {journal} {\bibinfo  {journal} {J. Chem. Phys.}\ }\textbf {\bibinfo {volume} {137}},\ \bibinfo {pages} {131102} (\bibinfo {year} {2012})}\BibitemShut {NoStop}%
\bibitem [{\citenamefont {Rinker}\ and\ \citenamefont {Steffen}(1977)}]{RinkerSteffen1977}%
  \BibitemOpen
  \bibfield  {author} {\bibinfo {author} {\bibfnamefont {G.~A.}\ \bibnamefont {Rinker}}\ and\ \bibinfo {author} {\bibfnamefont {R.~M.}\ \bibnamefont {Steffen}},\ }\href {https://doi.org/10.1016/0092-640X(77)90043-2} {\bibfield  {journal} {\bibinfo  {journal} {At. Data Nucl. Data Tables}\ }\textbf {\bibinfo {volume} {20}},\ \bibinfo {pages} {143} (\bibinfo {year} {1977})}\BibitemShut {NoStop}%
\bibitem [{\citenamefont {Steffen}(1981)}]{Steffen1981Muon}%
  \BibitemOpen
  \bibfield  {author} {\bibinfo {author} {\bibfnamefont {R.~M.}\ \bibnamefont {Steffen}},\ }\href {https://doi.org/10.1007/BF01020896} {\bibfield  {journal} {\bibinfo  {journal} {Hyperfine Interact.}\ }\textbf {\bibinfo {volume} {9}},\ \bibinfo {pages} {39} (\bibinfo {year} {1981})}\BibitemShut {NoStop}%
\bibitem [{\citenamefont {Steffen}(1985)}]{Steffen1985Muon}%
  \BibitemOpen
  \bibfield  {author} {\bibinfo {author} {\bibfnamefont {R.~M.}\ \bibnamefont {Steffen}},\ }\href {https://doi.org/10.1007/BF02354813} {\bibfield  {journal} {\bibinfo  {journal} {Hyperfine Interact.}\ }\textbf {\bibinfo {volume} {24}},\ \bibinfo {pages} {223} (\bibinfo {year} {1985})}\BibitemShut {NoStop}%
\bibitem [{\citenamefont {Michel}\ \emph {et~al.}(2017)\citenamefont {Michel}, \citenamefont {Oreshkina},\ and\ \citenamefont {Keitel}}]{Michel2017Muonic}%
  \BibitemOpen
  \bibfield  {author} {\bibinfo {author} {\bibfnamefont {N.}~\bibnamefont {Michel}}, \bibinfo {author} {\bibfnamefont {N.~S.}\ \bibnamefont {Oreshkina}},\ and\ \bibinfo {author} {\bibfnamefont {C.~H.}\ \bibnamefont {Keitel}},\ }\href {https://doi.org/10.1103/PhysRevA.96.032510} {\bibfield  {journal} {\bibinfo  {journal} {Phys. Rev. A}\ }\textbf {\bibinfo {volume} {96}},\ \bibinfo {pages} {032510} (\bibinfo {year} {2017})}\BibitemShut {NoStop}%
\bibitem [{\citenamefont {Dzuba}\ and\ \citenamefont {Flambaum}(2023)}]{DzubaFlambaum2023FNS}%
  \BibitemOpen
  \bibfield  {author} {\bibinfo {author} {\bibfnamefont {V.~A.}\ \bibnamefont {Dzuba}}\ and\ \bibinfo {author} {\bibfnamefont {V.~V.}\ \bibnamefont {Flambaum}},\ }\href@noop {} {\bibinfo {title} {Effect of finite nuclear size on the electric quadrupole hyperfine operator}} (\bibinfo {year} {2023}),\ \Eprint {https://arxiv.org/abs/2305.01208} {arXiv:2305.01208} \BibitemShut {NoStop}%
\bibitem [{\citenamefont {Jankovi{\'c}}\ \emph {et~al.}(2026)\citenamefont {Jankovi{\'c}}, \citenamefont {Hartmann}, \citenamefont {Bartel}, \citenamefont {Molique}, \citenamefont {Bonneau},\ and\ \citenamefont {Hervieux}}]{Jankovic2026Hyperfine}%
  \BibitemOpen
  \bibfield  {author} {\bibinfo {author} {\bibfnamefont {D.}~\bibnamefont {Jankovi{\'c}}}, \bibinfo {author} {\bibfnamefont {J.-G.}\ \bibnamefont {Hartmann}}, \bibinfo {author} {\bibfnamefont {J.}~\bibnamefont {Bartel}}, \bibinfo {author} {\bibfnamefont {H.}~\bibnamefont {Molique}}, \bibinfo {author} {\bibfnamefont {L.}~\bibnamefont {Bonneau}},\ and\ \bibinfo {author} {\bibfnamefont {P.-A.}\ \bibnamefont {Hervieux}},\ }\href@noop {} {\bibinfo {title} {The hyperfine interaction as a probe of the microscopic structure of the atomic nucleus}} (\bibinfo {year} {2026}),\ \Eprint {https://arxiv.org/abs/2602.18144} {arXiv:2602.18144} \BibitemShut {NoStop}%
\bibitem [{\citenamefont {Li}\ \emph {et~al.}(2021)\citenamefont {Li}, \citenamefont {Tang}, \citenamefont {Qiao},\ and\ \citenamefont {Shi}}]{Li2021RaPlus}%
  \BibitemOpen
  \bibfield  {author} {\bibinfo {author} {\bibfnamefont {F.-C.}\ \bibnamefont {Li}}, \bibinfo {author} {\bibfnamefont {Y.-B.}\ \bibnamefont {Tang}}, \bibinfo {author} {\bibfnamefont {H.-X.}\ \bibnamefont {Qiao}},\ and\ \bibinfo {author} {\bibfnamefont {T.-Y.}\ \bibnamefont {Shi}},\ }\href {https://doi.org/10.1088/1361-6455/ac1329} {\bibfield  {journal} {\bibinfo  {journal} {J. Phys. B: At. Mol. Opt. Phys.}\ }\textbf {\bibinfo {volume} {54}},\ \bibinfo {pages} {145004} (\bibinfo {year} {2021})}\BibitemShut {NoStop}%
\bibitem [{\citenamefont {Biero{\'n}}\ and\ \citenamefont {Pyykk{\"o}}(2005)}]{BieronPyykko2005Ra}%
  \BibitemOpen
  \bibfield  {author} {\bibinfo {author} {\bibfnamefont {J.}~\bibnamefont {Biero{\'n}}}\ and\ \bibinfo {author} {\bibfnamefont {P.}~\bibnamefont {Pyykk{\"o}}},\ }\href {https://doi.org/10.1103/PhysRevA.71.032502} {\bibfield  {journal} {\bibinfo  {journal} {Phys. Rev. A}\ }\textbf {\bibinfo {volume} {71}},\ \bibinfo {pages} {032502} (\bibinfo {year} {2005})}\BibitemShut {NoStop}%
\bibitem [{\citenamefont {Nezosi}\ \emph {et~al.}(2026)\citenamefont {Nezosi}, \citenamefont {Palmeri},\ and\ \citenamefont {J{\"o}nsson}}]{Nezosi2026RaHFS}%
  \BibitemOpen
  \bibfield  {author} {\bibinfo {author} {\bibfnamefont {L.}~\bibnamefont {Nezosi}}, \bibinfo {author} {\bibfnamefont {P.}~\bibnamefont {Palmeri}},\ and\ \bibinfo {author} {\bibfnamefont {P.}~\bibnamefont {J{\"o}nsson}},\ }\href {https://doi.org/10.3390/atoms14090072} {\bibfield  {journal} {\bibinfo  {journal} {Atoms}\ }\textbf {\bibinfo {volume} {14}},\ \bibinfo {pages} {72} (\bibinfo {year} {2026})}\BibitemShut {NoStop}%
\bibitem [{\citenamefont {Brown}\ and\ \citenamefont {Carrington}(2003)}]{BrownCarrington2003}%
  \BibitemOpen
  \bibfield  {author} {\bibinfo {author} {\bibfnamefont {J.~M.}\ \bibnamefont {Brown}}\ and\ \bibinfo {author} {\bibfnamefont {A.}~\bibnamefont {Carrington}},\ }\href {https://doi.org/10.1017/CBO9780511814808} {\emph {\bibinfo {title} {Rotational Spectroscopy of Diatomic Molecules}}}\ (\bibinfo  {publisher} {Cambridge University Press},\ \bibinfo {address} {Cambridge, UK},\ \bibinfo {year} {2003})\BibitemShut {NoStop}%
\bibitem [{\citenamefont {Tang}(2026)}]{Tang2026MgSrBa}%
  \BibitemOpen
  \bibfield  {author} {\bibinfo {author} {\bibfnamefont {Y.-B.}\ \bibnamefont {Tang}},\ }\href {https://doi.org/10.1103/1jxh-c6hw} {\bibfield  {journal} {\bibinfo  {journal} {Phys. Rev. A}\ }\textbf {\bibinfo {volume} {113}},\ \bibinfo {pages} {022812} (\bibinfo {year} {2026})}\BibitemShut {NoStop}%
\bibitem [{\citenamefont {Visscher}\ and\ \citenamefont {Dyall}(1997)}]{VisscherDyall1997}%
  \BibitemOpen
  \bibfield  {author} {\bibinfo {author} {\bibfnamefont {L.}~\bibnamefont {Visscher}}\ and\ \bibinfo {author} {\bibfnamefont {K.~G.}\ \bibnamefont {Dyall}},\ }\href {https://doi.org/10.1006/adnd.1997.0751} {\bibfield  {journal} {\bibinfo  {journal} {At. Data Nucl. Data Tables}\ }\textbf {\bibinfo {volume} {67}},\ \bibinfo {pages} {207} (\bibinfo {year} {1997})}\BibitemShut {NoStop}%
\bibitem [{\citenamefont {Visscher}\ \emph {et~al.}(1996)\citenamefont {Visscher}, \citenamefont {Lee},\ and\ \citenamefont {Dyall}}]{Visscher:96a}%
  \BibitemOpen
  \bibfield  {author} {\bibinfo {author} {\bibfnamefont {L.}~\bibnamefont {Visscher}}, \bibinfo {author} {\bibfnamefont {T.~J.}\ \bibnamefont {Lee}},\ and\ \bibinfo {author} {\bibfnamefont {K.~G.}\ \bibnamefont {Dyall}},\ }\href {https://doi.org/10.1063/1.472655} {\bibfield  {journal} {\bibinfo  {journal} {J. Chem. Phys.}\ }\textbf {\bibinfo {volume} {105}},\ \bibinfo {pages} {8769} (\bibinfo {year} {1996})}\BibitemShut {NoStop}%
\bibitem [{\citenamefont {Dyall}(2009)}]{Dyall:2009}%
  \BibitemOpen
  \bibfield  {author} {\bibinfo {author} {\bibfnamefont {K.~G.}\ \bibnamefont {Dyall}},\ }\href {https://doi.org/10.1021/jp905057q} {\bibfield  {journal} {\bibinfo  {journal} {J. Phys. Chem. A}\ }\textbf {\bibinfo {volume} {113}},\ \bibinfo {pages} {12638} (\bibinfo {year} {2009})}\BibitemShut {NoStop}%
\bibitem [{\citenamefont {Skripnikov}\ \emph {et~al.}(2017)\citenamefont {Skripnikov}, \citenamefont {Maison},\ and\ \citenamefont {Mosyagin}}]{Skripnikov:17a}%
  \BibitemOpen
  \bibfield  {author} {\bibinfo {author} {\bibfnamefont {L.~V.}\ \bibnamefont {Skripnikov}}, \bibinfo {author} {\bibfnamefont {D.~E.}\ \bibnamefont {Maison}},\ and\ \bibinfo {author} {\bibfnamefont {N.~S.}\ \bibnamefont {Mosyagin}},\ }\href {https://doi.org/10.1103/PhysRevA.95.022507} {\bibfield  {journal} {\bibinfo  {journal} {Phys. Rev. A}\ }\textbf {\bibinfo {volume} {95}},\ \bibinfo {pages} {022507} (\bibinfo {year} {2017})}\BibitemShut {NoStop}%
\bibitem [{\citenamefont {Dyall}(2016)}]{Dyall:2016}%
  \BibitemOpen
  \bibfield  {author} {\bibinfo {author} {\bibfnamefont {K.~G.}\ \bibnamefont {Dyall}},\ }\href {https://doi.org/10.1007/s00214-016-1884-y} {\bibfield  {journal} {\bibinfo  {journal} {Theor. Chem. Acc.}\ }\textbf {\bibinfo {volume} {135}},\ \bibinfo {pages} {128} (\bibinfo {year} {2016})}\BibitemShut {NoStop}%
\bibitem [{\citenamefont {Dunning}(1989)}]{Dunning:89}%
  \BibitemOpen
  \bibfield  {author} {\bibinfo {author} {\bibfnamefont {T.~H.}\ \bibnamefont {Dunning}, \bibfnamefont {Jr.}},\ }\href {https://doi.org/10.1063/1.456153} {\bibfield  {journal} {\bibinfo  {journal} {J.\ Chem.\ Phys.}\ }\textbf {\bibinfo {volume} {90}},\ \bibinfo {pages} {1007} (\bibinfo {year} {1989})}\BibitemShut {NoStop}%
\bibitem [{\citenamefont {K\'{a}llay}\ and\ \citenamefont {Surj\'{a}n}(2001)}]{Kallay:1}%
  \BibitemOpen
  \bibfield  {author} {\bibinfo {author} {\bibfnamefont {M.}~\bibnamefont {K\'{a}llay}}\ and\ \bibinfo {author} {\bibfnamefont {P.~R.}\ \bibnamefont {Surj\'{a}n}},\ }\href {https://doi.org/10.1063/1.1383290} {\bibfield  {journal} {\bibinfo  {journal} {J. Chem. Phys.}\ }\textbf {\bibinfo {volume} {115}},\ \bibinfo {pages} {2945} (\bibinfo {year} {2001})}\BibitemShut {NoStop}%
\bibitem [{\citenamefont {K\'{a}llay}\ and\ \citenamefont {Gauss}(2005)}]{Kallay:6}%
  \BibitemOpen
  \bibfield  {author} {\bibinfo {author} {\bibfnamefont {M.}~\bibnamefont {K\'{a}llay}}\ and\ \bibinfo {author} {\bibfnamefont {J.}~\bibnamefont {Gauss}},\ }\href {https://doi.org/10.1063/1.2121589} {\bibfield  {journal} {\bibinfo  {journal} {J. Chem. Phys.}\ }\textbf {\bibinfo {volume} {123}},\ \bibinfo {eid} {214105} (\bibinfo {year} {2005})}\BibitemShut {NoStop}%
\bibitem [{\citenamefont {Angeli}\ and\ \citenamefont {Marinova}(2013)}]{ANGELI201369}%
  \BibitemOpen
  \bibfield  {author} {\bibinfo {author} {\bibfnamefont {I.}~\bibnamefont {Angeli}}\ and\ \bibinfo {author} {\bibfnamefont {K.}~\bibnamefont {Marinova}},\ }\href {https://doi.org/10.1016/j.adt.2011.12.006} {\bibfield  {journal} {\bibinfo  {journal} {At. Data Nucl. Data Tables}\ }\textbf {\bibinfo {volume} {99}},\ \bibinfo {pages} {69} (\bibinfo {year} {2013})}\BibitemShut {NoStop}%
\bibitem [{\citenamefont {Shabaev}\ \emph {et~al.}(2013)\citenamefont {Shabaev}, \citenamefont {Tupitsyn},\ and\ \citenamefont {Yerokhin}}]{Shabaev:13}%
  \BibitemOpen
  \bibfield  {author} {\bibinfo {author} {\bibfnamefont {V.~M.}\ \bibnamefont {Shabaev}}, \bibinfo {author} {\bibfnamefont {I.~I.}\ \bibnamefont {Tupitsyn}},\ and\ \bibinfo {author} {\bibfnamefont {V.~A.}\ \bibnamefont {Yerokhin}},\ }\href {https://doi.org/10.1103/PhysRevA.88.012513} {\bibfield  {journal} {\bibinfo  {journal} {Phys.\ Rev.\ A}\ }\textbf {\bibinfo {volume} {88}},\ \bibinfo {pages} {012513} (\bibinfo {year} {2013})}\BibitemShut {NoStop}%
\bibitem [{\citenamefont {Skripnikov}(2021)}]{Skripnikov:2021a}%
  \BibitemOpen
  \bibfield  {author} {\bibinfo {author} {\bibfnamefont {L.~V.}\ \bibnamefont {Skripnikov}},\ }\href {https://doi.org/10.1063/5.0053659} {\bibfield  {journal} {\bibinfo  {journal} {J. Chem. Phys.}\ }\textbf {\bibinfo {volume} {154}},\ \bibinfo {pages} {201101} (\bibinfo {year} {2021})}\BibitemShut {NoStop}%
\bibitem [{\citenamefont {Visscher}\ \emph {et~al.}(2017)\citenamefont {Visscher}, \citenamefont {Jensen}, \citenamefont {Bast},\ and\ \citenamefont {Saue}}]{DIRAC17}%
  \BibitemOpen
  \bibfield  {author} {\bibinfo {author} {\bibfnamefont {L.}~\bibnamefont {Visscher}}, \bibinfo {author} {\bibfnamefont {H.~J.~A.}\ \bibnamefont {Jensen}}, \bibinfo {author} {\bibfnamefont {R.}~\bibnamefont {Bast}},\ and\ \bibinfo {author} {\bibfnamefont {T.}~\bibnamefont {Saue}},\ }\href {https://www.diracprogram.org/doku.php?id=citation#dirac17} {\bibinfo {title} {{DIRAC}, a relativistic ab initio electronic structure program, release {DIRAC17}}} (\bibinfo {year} {2017}),\ \bibinfo {note} {software release}\BibitemShut {NoStop}%
\bibitem [{\citenamefont {Lebedev}(1977)}]{Lebedev1977}%
  \BibitemOpen
  \bibfield  {author} {\bibinfo {author} {\bibfnamefont {V.~I.}\ \bibnamefont {Lebedev}},\ }\href {https://doi.org/10.1007/BF00966954} {\bibfield  {journal} {\bibinfo  {journal} {Sib. Math. J.}\ }\textbf {\bibinfo {volume} {18}},\ \bibinfo {pages} {99} (\bibinfo {year} {1977})}\BibitemShut {NoStop}%
\bibitem [{\citenamefont {Lebedev}\ and\ \citenamefont {Laikov}(1999)}]{LebedevLaikov1999}%
  \BibitemOpen
  \bibfield  {author} {\bibinfo {author} {\bibfnamefont {V.~I.}\ \bibnamefont {Lebedev}}\ and\ \bibinfo {author} {\bibfnamefont {D.~N.}\ \bibnamefont {Laikov}},\ }\href@noop {} {\bibfield  {journal} {\bibinfo  {journal} {Dokl. Math.}\ }\textbf {\bibinfo {volume} {59}},\ \bibinfo {pages} {477} (\bibinfo {year} {1999})}\BibitemShut {NoStop}%
\bibitem [{\citenamefont {Saue}\ \emph {et~al.}(2020)\citenamefont {Saue}, \citenamefont {Bast}, \citenamefont {Gomes}, \citenamefont {Jensen}, \citenamefont {Visscher}, \citenamefont {Aucar}, \citenamefont {Di~Remigio}, \citenamefont {Dyall}, \citenamefont {Eliav}, \citenamefont {Fasshauer}, \citenamefont {Fleig}, \citenamefont {Halbert}, \citenamefont {Hedeg{\r{a}}rd}, \citenamefont {Helmich-Paris}, \citenamefont {Ilia{\v{s}}}, \citenamefont {Jacob}, \citenamefont {Knecht}, \citenamefont {Laerdahl}, \citenamefont {Vidal}, \citenamefont {Nayak}, \citenamefont {Olejniczak}, \citenamefont {Olsen}, \citenamefont {Pernpointner}, \citenamefont {Senjean}, \citenamefont {Shee}, \citenamefont {Sunaga},\ and\ \citenamefont {van Stralen}}]{Saue:2020}%
  \BibitemOpen
  \bibfield  {author} {\bibinfo {author} {\bibfnamefont {T.}~\bibnamefont {Saue}}, \bibinfo {author} {\bibfnamefont {R.}~\bibnamefont {Bast}}, \bibinfo {author} {\bibfnamefont {A.~S.~P.}\ \bibnamefont {Gomes}}, \bibinfo {author} {\bibfnamefont {H.~J.~A.}\ \bibnamefont {Jensen}}, \bibinfo {author} {\bibfnamefont {L.}~\bibnamefont {Visscher}}, \bibinfo {author} {\bibfnamefont {I.~A.}\ \bibnamefont {Aucar}}, \bibinfo {author} {\bibfnamefont {R.}~\bibnamefont {Di~Remigio}}, \bibinfo {author} {\bibfnamefont {K.~G.}\ \bibnamefont {Dyall}}, \bibinfo {author} {\bibfnamefont {E.}~\bibnamefont {Eliav}}, \bibinfo {author} {\bibfnamefont {E.}~\bibnamefont {Fasshauer}}, \bibinfo {author} {\bibfnamefont {T.}~\bibnamefont {Fleig}}, \bibinfo {author} {\bibfnamefont {L.}~\bibnamefont {Halbert}}, \bibinfo {author} {\bibfnamefont {E.~D.}\ \bibnamefont {Hedeg{\r{a}}rd}}, \bibinfo {author} {\bibfnamefont {B.}~\bibnamefont {Helmich-Paris}}, \bibinfo {author} {\bibfnamefont {M.}~\bibnamefont {Ilia{\v{s}}}}, \bibinfo {author}
  {\bibfnamefont {C.~R.}\ \bibnamefont {Jacob}}, \bibinfo {author} {\bibfnamefont {S.}~\bibnamefont {Knecht}}, \bibinfo {author} {\bibfnamefont {J.~K.}\ \bibnamefont {Laerdahl}}, \bibinfo {author} {\bibfnamefont {M.~L.}\ \bibnamefont {Vidal}}, \bibinfo {author} {\bibfnamefont {M.~K.}\ \bibnamefont {Nayak}}, \bibinfo {author} {\bibfnamefont {M.}~\bibnamefont {Olejniczak}}, \bibinfo {author} {\bibfnamefont {J.~M.~H.}\ \bibnamefont {Olsen}}, \bibinfo {author} {\bibfnamefont {M.}~\bibnamefont {Pernpointner}}, \bibinfo {author} {\bibfnamefont {B.}~\bibnamefont {Senjean}}, \bibinfo {author} {\bibfnamefont {A.}~\bibnamefont {Shee}}, \bibinfo {author} {\bibfnamefont {A.}~\bibnamefont {Sunaga}},\ and\ \bibinfo {author} {\bibfnamefont {J.~N.~P.}\ \bibnamefont {van Stralen}},\ }\href {https://doi.org/10.1063/5.0004844} {\bibfield  {journal} {\bibinfo  {journal} {J. Chem. Phys.}\ }\textbf {\bibinfo {volume} {152}},\ \bibinfo {pages} {204104} (\bibinfo {year} {2020})}\BibitemShut {NoStop}%
\bibitem [{\citenamefont {K{\'{a}}llay}\ \emph {et~al.}(2020)\citenamefont {K{\'{a}}llay}, \citenamefont {Nagy}, \citenamefont {Mester}, \citenamefont {Rolik}, \citenamefont {Samu}, \citenamefont {Csontos}, \citenamefont {Cs{\'{o}}ka}, \citenamefont {Szab{\'{o}}}, \citenamefont {Gyevi-Nagy}, \citenamefont {H{\'{e}}gely}, \citenamefont {Ladj{\'{a}}nszki}, \citenamefont {Szegedy}, \citenamefont {Lad{\'{o}}czki}, \citenamefont {Petrov}, \citenamefont {Farkas}, \citenamefont {Mezei},\ and\ \citenamefont {Ganyecz}}]{MRCC2020}%
  \BibitemOpen
  \bibfield  {author} {\bibinfo {author} {\bibfnamefont {M.}~\bibnamefont {K{\'{a}}llay}}, \bibinfo {author} {\bibfnamefont {P.~R.}\ \bibnamefont {Nagy}}, \bibinfo {author} {\bibfnamefont {D.}~\bibnamefont {Mester}}, \bibinfo {author} {\bibfnamefont {Z.}~\bibnamefont {Rolik}}, \bibinfo {author} {\bibfnamefont {G.}~\bibnamefont {Samu}}, \bibinfo {author} {\bibfnamefont {J.}~\bibnamefont {Csontos}}, \bibinfo {author} {\bibfnamefont {J.}~\bibnamefont {Cs{\'{o}}ka}}, \bibinfo {author} {\bibfnamefont {P.~B.}\ \bibnamefont {Szab{\'{o}}}}, \bibinfo {author} {\bibfnamefont {L.}~\bibnamefont {Gyevi-Nagy}}, \bibinfo {author} {\bibfnamefont {B.}~\bibnamefont {H{\'{e}}gely}}, \bibinfo {author} {\bibfnamefont {I.}~\bibnamefont {Ladj{\'{a}}nszki}}, \bibinfo {author} {\bibfnamefont {L.}~\bibnamefont {Szegedy}}, \bibinfo {author} {\bibfnamefont {B.}~\bibnamefont {Lad{\'{o}}czki}}, \bibinfo {author} {\bibfnamefont {K.}~\bibnamefont {Petrov}}, \bibinfo {author} {\bibfnamefont {M.}~\bibnamefont {Farkas}}, \bibinfo {author}
  {\bibfnamefont {P.~D.}\ \bibnamefont {Mezei}},\ and\ \bibinfo {author} {\bibfnamefont {{\'{A}}.}~\bibnamefont {Ganyecz}},\ }\href {https://doi.org/10.1063/1.5142048} {\bibfield  {journal} {\bibinfo  {journal} {J. Chem. Phys.}\ }\textbf {\bibinfo {volume} {152}},\ \bibinfo {pages} {074107} (\bibinfo {year} {2020})}\BibitemShut {NoStop}%
\bibitem [{\citenamefont {Sun}\ \emph {et~al.}(2020)\citenamefont {Sun}, \citenamefont {Zhang}, \citenamefont {Banerjee}, \citenamefont {Bao}, \citenamefont {Barbry}, \citenamefont {Blunt}, \citenamefont {Bogdanov}, \citenamefont {Booth}, \citenamefont {Chen}, \citenamefont {Cui}, \citenamefont {Eriksen}, \citenamefont {Gao}, \citenamefont {Guo}, \citenamefont {Hermann}, \citenamefont {Hermes}, \citenamefont {Koh}, \citenamefont {Koval}, \citenamefont {Lehtola}, \citenamefont {Li}, \citenamefont {Liu}, \citenamefont {Mardirossian}, \citenamefont {McClain}, \citenamefont {Motta}, \citenamefont {Mussard}, \citenamefont {Pham}, \citenamefont {Pulkin}, \citenamefont {Purwanto}, \citenamefont {Robinson}, \citenamefont {Ronca}, \citenamefont {Sayfutyarova}, \citenamefont {Scheurer}, \citenamefont {Schurkus}, \citenamefont {Smith}, \citenamefont {Sun}, \citenamefont {Sun}, \citenamefont {Upadhyay}, \citenamefont {Wagner}, \citenamefont {Wang}, \citenamefont {White}, \citenamefont {Whitfield}, \citenamefont
  {Williamson}, \citenamefont {Wouters}, \citenamefont {Yang}, \citenamefont {Yu}, \citenamefont {Zhu}, \citenamefont {Berkelbach}, \citenamefont {Sharma}, \citenamefont {Sokolov},\ and\ \citenamefont {Chan}}]{Sun2020PySCF}%
  \BibitemOpen
  \bibfield  {author} {\bibinfo {author} {\bibfnamefont {Q.}~\bibnamefont {Sun}}, \bibinfo {author} {\bibfnamefont {X.}~\bibnamefont {Zhang}}, \bibinfo {author} {\bibfnamefont {S.}~\bibnamefont {Banerjee}}, \bibinfo {author} {\bibfnamefont {P.}~\bibnamefont {Bao}}, \bibinfo {author} {\bibfnamefont {M.}~\bibnamefont {Barbry}}, \bibinfo {author} {\bibfnamefont {N.~S.}\ \bibnamefont {Blunt}}, \bibinfo {author} {\bibfnamefont {N.~A.}\ \bibnamefont {Bogdanov}}, \bibinfo {author} {\bibfnamefont {G.~H.}\ \bibnamefont {Booth}}, \bibinfo {author} {\bibfnamefont {J.}~\bibnamefont {Chen}}, \bibinfo {author} {\bibfnamefont {Z.-H.}\ \bibnamefont {Cui}}, \bibinfo {author} {\bibfnamefont {J.~J.}\ \bibnamefont {Eriksen}}, \bibinfo {author} {\bibfnamefont {Y.}~\bibnamefont {Gao}}, \bibinfo {author} {\bibfnamefont {S.}~\bibnamefont {Guo}}, \bibinfo {author} {\bibfnamefont {J.}~\bibnamefont {Hermann}}, \bibinfo {author} {\bibfnamefont {M.~R.}\ \bibnamefont {Hermes}}, \bibinfo {author} {\bibfnamefont {K.}~\bibnamefont {Koh}},
  \bibinfo {author} {\bibfnamefont {P.}~\bibnamefont {Koval}}, \bibinfo {author} {\bibfnamefont {S.}~\bibnamefont {Lehtola}}, \bibinfo {author} {\bibfnamefont {Z.}~\bibnamefont {Li}}, \bibinfo {author} {\bibfnamefont {J.}~\bibnamefont {Liu}}, \bibinfo {author} {\bibfnamefont {N.}~\bibnamefont {Mardirossian}}, \bibinfo {author} {\bibfnamefont {J.~D.}\ \bibnamefont {McClain}}, \bibinfo {author} {\bibfnamefont {M.}~\bibnamefont {Motta}}, \bibinfo {author} {\bibfnamefont {B.}~\bibnamefont {Mussard}}, \bibinfo {author} {\bibfnamefont {H.~Q.}\ \bibnamefont {Pham}}, \bibinfo {author} {\bibfnamefont {A.}~\bibnamefont {Pulkin}}, \bibinfo {author} {\bibfnamefont {W.}~\bibnamefont {Purwanto}}, \bibinfo {author} {\bibfnamefont {P.~J.}\ \bibnamefont {Robinson}}, \bibinfo {author} {\bibfnamefont {E.}~\bibnamefont {Ronca}}, \bibinfo {author} {\bibfnamefont {E.~R.}\ \bibnamefont {Sayfutyarova}}, \bibinfo {author} {\bibfnamefont {M.}~\bibnamefont {Scheurer}}, \bibinfo {author} {\bibfnamefont {H.~F.}\ \bibnamefont {Schurkus}},
  \bibinfo {author} {\bibfnamefont {J.~E.~T.}\ \bibnamefont {Smith}}, \bibinfo {author} {\bibfnamefont {C.}~\bibnamefont {Sun}}, \bibinfo {author} {\bibfnamefont {S.-N.}\ \bibnamefont {Sun}}, \bibinfo {author} {\bibfnamefont {S.}~\bibnamefont {Upadhyay}}, \bibinfo {author} {\bibfnamefont {L.~K.}\ \bibnamefont {Wagner}}, \bibinfo {author} {\bibfnamefont {X.}~\bibnamefont {Wang}}, \bibinfo {author} {\bibfnamefont {A.}~\bibnamefont {White}}, \bibinfo {author} {\bibfnamefont {J.~D.}\ \bibnamefont {Whitfield}}, \bibinfo {author} {\bibfnamefont {M.~J.}\ \bibnamefont {Williamson}}, \bibinfo {author} {\bibfnamefont {S.}~\bibnamefont {Wouters}}, \bibinfo {author} {\bibfnamefont {J.}~\bibnamefont {Yang}}, \bibinfo {author} {\bibfnamefont {J.~M.}\ \bibnamefont {Yu}}, \bibinfo {author} {\bibfnamefont {T.}~\bibnamefont {Zhu}}, \bibinfo {author} {\bibfnamefont {T.~C.}\ \bibnamefont {Berkelbach}}, \bibinfo {author} {\bibfnamefont {S.}~\bibnamefont {Sharma}}, \bibinfo {author} {\bibfnamefont {A.~Y.}\ \bibnamefont
  {Sokolov}},\ and\ \bibinfo {author} {\bibfnamefont {G.~K.-L.}\ \bibnamefont {Chan}},\ }\href {https://doi.org/10.1063/5.0006074} {\bibfield  {journal} {\bibinfo  {journal} {J. Chem. Phys.}\ }\textbf {\bibinfo {volume} {153}},\ \bibinfo {pages} {024109} (\bibinfo {year} {2020})}\BibitemShut {NoStop}%
\bibitem [{\citenamefont {Li~Manni}\ \emph {et~al.}(2023)\citenamefont {Li~Manni}, \citenamefont {Fdez.~Galv{\'{a}}n}, \citenamefont {Alavi}, \citenamefont {Aleotti}, \citenamefont {Aquilante}, \citenamefont {Autschbach}, \citenamefont {Avagliano}, \citenamefont {Baiardi}, \citenamefont {Bao}, \citenamefont {Battaglia}, \citenamefont {Birnoschi}, \citenamefont {Blanco-Gonz{\'{a}}lez}, \citenamefont {Bokarev}, \citenamefont {Broer}, \citenamefont {Cacciari}, \citenamefont {Calio}, \citenamefont {Carlson}, \citenamefont {Carvalho~Couto}, \citenamefont {Cerd{\'{a}}n}, \citenamefont {Chibotaru}, \citenamefont {Chilton}, \citenamefont {Church}, \citenamefont {Conti}, \citenamefont {Coriani}, \citenamefont {Cu{\'{e}}llar-Zuquin}, \citenamefont {Daoud}, \citenamefont {Dattani}, \citenamefont {Decleva}, \citenamefont {de~Graaf}, \citenamefont {Delcey}, \citenamefont {De~Vico}, \citenamefont {Dobrautz}, \citenamefont {Dong}, \citenamefont {Feng}, \citenamefont {Ferr{\'{e}}}, \citenamefont {Filatov(Gulak)},
  \citenamefont {Gagliardi}, \citenamefont {Garavelli}, \citenamefont {Gonz{\'{a}}lez}, \citenamefont {Guan}, \citenamefont {Guo}, \citenamefont {Hennefarth}, \citenamefont {Hermes}, \citenamefont {Hoyer}, \citenamefont {Huix-Rotllant}, \citenamefont {Jaiswal}, \citenamefont {Kaiser}, \citenamefont {Kaliakin}, \citenamefont {Khamesian}, \citenamefont {King}, \citenamefont {Kochetov}, \citenamefont {Kro{\'{s}}nicki}, \citenamefont {Kumaar}, \citenamefont {Larsson}, \citenamefont {Lehtola}, \citenamefont {Lepetit}, \citenamefont {Lischka}, \citenamefont {L{\'{o}}pez~R{\'{i}}os}, \citenamefont {Lundberg}, \citenamefont {Ma}, \citenamefont {Mai}, \citenamefont {Marquetand}, \citenamefont {Merritt}, \citenamefont {Montorsi}, \citenamefont {M{\"{o}}rchen}, \citenamefont {Nenov}, \citenamefont {Nguyen}, \citenamefont {Nishimoto}, \citenamefont {Oakley}, \citenamefont {Olivucci}, \citenamefont {Oppel}, \citenamefont {Padula}, \citenamefont {Pandharkar}, \citenamefont {Phung}, \citenamefont {Plasser}, \citenamefont
  {Raggi}, \citenamefont {Rebolini}, \citenamefont {Reiher}, \citenamefont {Rivalta}, \citenamefont {Roca-Sanju{\'{a}}n}, \citenamefont {Romig}, \citenamefont {Safari}, \citenamefont {S{\'{a}}nchez-Mansilla}, \citenamefont {Sand}, \citenamefont {Schapiro}, \citenamefont {Scott}, \citenamefont {Segarra-Mart{\'{i}}}, \citenamefont {Segatta}, \citenamefont {Sergentu}, \citenamefont {Sharma}, \citenamefont {Shepard}, \citenamefont {Shu}, \citenamefont {Staab}, \citenamefont {Straatsma}, \citenamefont {S{\o}rensen}, \citenamefont {Tenorio}, \citenamefont {Truhlar}, \citenamefont {Ungur}, \citenamefont {Vacher}, \citenamefont {Veryazov}, \citenamefont {Vo{\ss}}, \citenamefont {Weser}, \citenamefont {Wu}, \citenamefont {Yang}, \citenamefont {Yarkony}, \citenamefont {Zhou}, \citenamefont {Zobel},\ and\ \citenamefont {Lindh}}]{LiManni2023OpenMolcas}%
  \BibitemOpen
  \bibfield  {author} {\bibinfo {author} {\bibfnamefont {G.}~\bibnamefont {Li~Manni}}, \bibinfo {author} {\bibfnamefont {I.}~\bibnamefont {Fdez.~Galv{\'{a}}n}}, \bibinfo {author} {\bibfnamefont {A.}~\bibnamefont {Alavi}}, \bibinfo {author} {\bibfnamefont {F.}~\bibnamefont {Aleotti}}, \bibinfo {author} {\bibfnamefont {F.}~\bibnamefont {Aquilante}}, \bibinfo {author} {\bibfnamefont {J.}~\bibnamefont {Autschbach}}, \bibinfo {author} {\bibfnamefont {D.}~\bibnamefont {Avagliano}}, \bibinfo {author} {\bibfnamefont {A.}~\bibnamefont {Baiardi}}, \bibinfo {author} {\bibfnamefont {J.~J.}\ \bibnamefont {Bao}}, \bibinfo {author} {\bibfnamefont {S.}~\bibnamefont {Battaglia}}, \bibinfo {author} {\bibfnamefont {L.}~\bibnamefont {Birnoschi}}, \bibinfo {author} {\bibfnamefont {A.}~\bibnamefont {Blanco-Gonz{\'{a}}lez}}, \bibinfo {author} {\bibfnamefont {S.~I.}\ \bibnamefont {Bokarev}}, \bibinfo {author} {\bibfnamefont {R.}~\bibnamefont {Broer}}, \bibinfo {author} {\bibfnamefont {R.}~\bibnamefont {Cacciari}}, \bibinfo {author}
  {\bibfnamefont {P.~B.}\ \bibnamefont {Calio}}, \bibinfo {author} {\bibfnamefont {R.~K.}\ \bibnamefont {Carlson}}, \bibinfo {author} {\bibfnamefont {R.}~\bibnamefont {Carvalho~Couto}}, \bibinfo {author} {\bibfnamefont {L.}~\bibnamefont {Cerd{\'{a}}n}}, \bibinfo {author} {\bibfnamefont {L.~F.}\ \bibnamefont {Chibotaru}}, \bibinfo {author} {\bibfnamefont {N.~F.}\ \bibnamefont {Chilton}}, \bibinfo {author} {\bibfnamefont {J.~R.}\ \bibnamefont {Church}}, \bibinfo {author} {\bibfnamefont {I.}~\bibnamefont {Conti}}, \bibinfo {author} {\bibfnamefont {S.}~\bibnamefont {Coriani}}, \bibinfo {author} {\bibfnamefont {J.}~\bibnamefont {Cu{\'{e}}llar-Zuquin}}, \bibinfo {author} {\bibfnamefont {R.~E.}\ \bibnamefont {Daoud}}, \bibinfo {author} {\bibfnamefont {N.}~\bibnamefont {Dattani}}, \bibinfo {author} {\bibfnamefont {P.}~\bibnamefont {Decleva}}, \bibinfo {author} {\bibfnamefont {C.}~\bibnamefont {de~Graaf}}, \bibinfo {author} {\bibfnamefont {M.~G.}\ \bibnamefont {Delcey}}, \bibinfo {author} {\bibfnamefont
  {L.}~\bibnamefont {De~Vico}}, \bibinfo {author} {\bibfnamefont {W.}~\bibnamefont {Dobrautz}}, \bibinfo {author} {\bibfnamefont {S.~S.}\ \bibnamefont {Dong}}, \bibinfo {author} {\bibfnamefont {R.}~\bibnamefont {Feng}}, \bibinfo {author} {\bibfnamefont {N.}~\bibnamefont {Ferr{\'{e}}}}, \bibinfo {author} {\bibfnamefont {M.}~\bibnamefont {Filatov(Gulak)}}, \bibinfo {author} {\bibfnamefont {L.}~\bibnamefont {Gagliardi}}, \bibinfo {author} {\bibfnamefont {M.}~\bibnamefont {Garavelli}}, \bibinfo {author} {\bibfnamefont {L.}~\bibnamefont {Gonz{\'{a}}lez}}, \bibinfo {author} {\bibfnamefont {Y.}~\bibnamefont {Guan}}, \bibinfo {author} {\bibfnamefont {M.}~\bibnamefont {Guo}}, \bibinfo {author} {\bibfnamefont {M.~R.}\ \bibnamefont {Hennefarth}}, \bibinfo {author} {\bibfnamefont {M.~R.}\ \bibnamefont {Hermes}}, \bibinfo {author} {\bibfnamefont {C.~E.}\ \bibnamefont {Hoyer}}, \bibinfo {author} {\bibfnamefont {M.}~\bibnamefont {Huix-Rotllant}}, \bibinfo {author} {\bibfnamefont {V.~K.}\ \bibnamefont {Jaiswal}}, \bibinfo
  {author} {\bibfnamefont {A.}~\bibnamefont {Kaiser}}, \bibinfo {author} {\bibfnamefont {D.~S.}\ \bibnamefont {Kaliakin}}, \bibinfo {author} {\bibfnamefont {M.}~\bibnamefont {Khamesian}}, \bibinfo {author} {\bibfnamefont {D.~S.}\ \bibnamefont {King}}, \bibinfo {author} {\bibfnamefont {V.}~\bibnamefont {Kochetov}}, \bibinfo {author} {\bibfnamefont {M.}~\bibnamefont {Kro{\'{s}}nicki}}, \bibinfo {author} {\bibfnamefont {A.~A.}\ \bibnamefont {Kumaar}}, \bibinfo {author} {\bibfnamefont {E.~D.}\ \bibnamefont {Larsson}}, \bibinfo {author} {\bibfnamefont {S.}~\bibnamefont {Lehtola}}, \bibinfo {author} {\bibfnamefont {M.-B.}\ \bibnamefont {Lepetit}}, \bibinfo {author} {\bibfnamefont {H.}~\bibnamefont {Lischka}}, \bibinfo {author} {\bibfnamefont {P.}~\bibnamefont {L{\'{o}}pez~R{\'{i}}os}}, \bibinfo {author} {\bibfnamefont {M.}~\bibnamefont {Lundberg}}, \bibinfo {author} {\bibfnamefont {D.}~\bibnamefont {Ma}}, \bibinfo {author} {\bibfnamefont {S.}~\bibnamefont {Mai}}, \bibinfo {author} {\bibfnamefont {P.}~\bibnamefont
  {Marquetand}}, \bibinfo {author} {\bibfnamefont {I.~C.~D.}\ \bibnamefont {Merritt}}, \bibinfo {author} {\bibfnamefont {F.}~\bibnamefont {Montorsi}}, \bibinfo {author} {\bibfnamefont {M.}~\bibnamefont {M{\"{o}}rchen}}, \bibinfo {author} {\bibfnamefont {A.}~\bibnamefont {Nenov}}, \bibinfo {author} {\bibfnamefont {V.~H.~A.}\ \bibnamefont {Nguyen}}, \bibinfo {author} {\bibfnamefont {Y.}~\bibnamefont {Nishimoto}}, \bibinfo {author} {\bibfnamefont {M.~S.}\ \bibnamefont {Oakley}}, \bibinfo {author} {\bibfnamefont {M.}~\bibnamefont {Olivucci}}, \bibinfo {author} {\bibfnamefont {M.}~\bibnamefont {Oppel}}, \bibinfo {author} {\bibfnamefont {D.}~\bibnamefont {Padula}}, \bibinfo {author} {\bibfnamefont {R.}~\bibnamefont {Pandharkar}}, \bibinfo {author} {\bibfnamefont {Q.~M.}\ \bibnamefont {Phung}}, \bibinfo {author} {\bibfnamefont {F.}~\bibnamefont {Plasser}}, \bibinfo {author} {\bibfnamefont {G.}~\bibnamefont {Raggi}}, \bibinfo {author} {\bibfnamefont {E.}~\bibnamefont {Rebolini}}, \bibinfo {author} {\bibfnamefont
  {M.}~\bibnamefont {Reiher}}, \bibinfo {author} {\bibfnamefont {I.}~\bibnamefont {Rivalta}}, \bibinfo {author} {\bibfnamefont {D.}~\bibnamefont {Roca-Sanju{\'{a}}n}}, \bibinfo {author} {\bibfnamefont {T.}~\bibnamefont {Romig}}, \bibinfo {author} {\bibfnamefont {A.~A.}\ \bibnamefont {Safari}}, \bibinfo {author} {\bibfnamefont {A.}~\bibnamefont {S{\'{a}}nchez-Mansilla}}, \bibinfo {author} {\bibfnamefont {A.~M.}\ \bibnamefont {Sand}}, \bibinfo {author} {\bibfnamefont {I.}~\bibnamefont {Schapiro}}, \bibinfo {author} {\bibfnamefont {T.~R.}\ \bibnamefont {Scott}}, \bibinfo {author} {\bibfnamefont {J.}~\bibnamefont {Segarra-Mart{\'{i}}}}, \bibinfo {author} {\bibfnamefont {F.}~\bibnamefont {Segatta}}, \bibinfo {author} {\bibfnamefont {D.-C.}\ \bibnamefont {Sergentu}}, \bibinfo {author} {\bibfnamefont {P.}~\bibnamefont {Sharma}}, \bibinfo {author} {\bibfnamefont {R.}~\bibnamefont {Shepard}}, \bibinfo {author} {\bibfnamefont {Y.}~\bibnamefont {Shu}}, \bibinfo {author} {\bibfnamefont {J.~K.}\ \bibnamefont {Staab}},
  \bibinfo {author} {\bibfnamefont {T.~P.}\ \bibnamefont {Straatsma}}, \bibinfo {author} {\bibfnamefont {L.~K.}\ \bibnamefont {S{\o}rensen}}, \bibinfo {author} {\bibfnamefont {B.~N.~C.}\ \bibnamefont {Tenorio}}, \bibinfo {author} {\bibfnamefont {D.~G.}\ \bibnamefont {Truhlar}}, \bibinfo {author} {\bibfnamefont {L.}~\bibnamefont {Ungur}}, \bibinfo {author} {\bibfnamefont {M.}~\bibnamefont {Vacher}}, \bibinfo {author} {\bibfnamefont {V.}~\bibnamefont {Veryazov}}, \bibinfo {author} {\bibfnamefont {T.~A.}\ \bibnamefont {Vo{\ss}}}, \bibinfo {author} {\bibfnamefont {O.}~\bibnamefont {Weser}}, \bibinfo {author} {\bibfnamefont {D.}~\bibnamefont {Wu}}, \bibinfo {author} {\bibfnamefont {X.}~\bibnamefont {Yang}}, \bibinfo {author} {\bibfnamefont {D.}~\bibnamefont {Yarkony}}, \bibinfo {author} {\bibfnamefont {C.}~\bibnamefont {Zhou}}, \bibinfo {author} {\bibfnamefont {J.~P.}\ \bibnamefont {Zobel}},\ and\ \bibinfo {author} {\bibfnamefont {R.}~\bibnamefont {Lindh}},\ }\href {https://doi.org/10.1021/acs.jctc.3c00182}
  {\bibfield  {journal} {\bibinfo  {journal} {J. Chem. Theory Comput.}\ }\textbf {\bibinfo {volume} {19}},\ \bibinfo {pages} {6933} (\bibinfo {year} {2023})}\BibitemShut {NoStop}%
\bibitem [{\citenamefont {Skripnikov}(2016)}]{Skripnikov:16b}%
  \BibitemOpen
  \bibfield  {author} {\bibinfo {author} {\bibfnamefont {L.~V.}\ \bibnamefont {Skripnikov}},\ }\href {https://doi.org/10.1063/1.4968229} {\bibfield  {journal} {\bibinfo  {journal} {J. Chem. Phys.}\ }\textbf {\bibinfo {volume} {145}},\ \bibinfo {pages} {214301} (\bibinfo {year} {2016})}\BibitemShut {NoStop}%
\bibitem [{\citenamefont {Fabbro}\ \emph {et~al.}(2026)\citenamefont {Fabbro}, \citenamefont {Brandejs},\ and\ \citenamefont {Saue}}]{Fabbro2026SrNQM}%
  \BibitemOpen
  \bibfield  {author} {\bibinfo {author} {\bibfnamefont {G.}~\bibnamefont {Fabbro}}, \bibinfo {author} {\bibfnamefont {J.}~\bibnamefont {Brandejs}},\ and\ \bibinfo {author} {\bibfnamefont {T.}~\bibnamefont {Saue}},\ }\href {https://doi.org/10.1021/acs.jpca.6c00683} {\bibfield  {journal} {\bibinfo  {journal} {J. Phys. Chem. A}\ }\textbf {\bibinfo {volume} {130}},\ \bibinfo {pages} {3187} (\bibinfo {year} {2026})}\BibitemShut {NoStop}%
\bibitem [{\citenamefont {Heully}\ and\ \citenamefont {M{\aa}rtensson-Pendrill}(1985)}]{Heully1985RaHFS}%
  \BibitemOpen
  \bibfield  {author} {\bibinfo {author} {\bibfnamefont {J.-L.}\ \bibnamefont {Heully}}\ and\ \bibinfo {author} {\bibfnamefont {A.-M.}\ \bibnamefont {M{\aa}rtensson-Pendrill}},\ }\href {https://doi.org/10.1088/0031-8949/31/3/003} {\bibfield  {journal} {\bibinfo  {journal} {Phys. Scr.}\ }\textbf {\bibinfo {volume} {31}},\ \bibinfo {pages} {169} (\bibinfo {year} {1985})}\BibitemShut {NoStop}%
\bibitem [{\citenamefont {Andriessen}\ \emph {et~al.}(1992)\citenamefont {Andriessen}, \citenamefont {Postma}, \citenamefont {van~den Brink},\ and\ \citenamefont {Das}}]{Andriessen1992RaHFS}%
  \BibitemOpen
  \bibfield  {author} {\bibinfo {author} {\bibfnamefont {J.}~\bibnamefont {Andriessen}}, \bibinfo {author} {\bibfnamefont {H.}~\bibnamefont {Postma}}, \bibinfo {author} {\bibfnamefont {A.~M.}\ \bibnamefont {van~den Brink}},\ and\ \bibinfo {author} {\bibfnamefont {T.~P.}\ \bibnamefont {Das}},\ }\href {https://doi.org/10.1103/PhysRevA.45.1389} {\bibfield  {journal} {\bibinfo  {journal} {Phys. Rev. A}\ }\textbf {\bibinfo {volume} {45}},\ \bibinfo {pages} {1389} (\bibinfo {year} {1992})}\BibitemShut {NoStop}%
\bibitem [{\citenamefont {Biero{\'n}}(2005)}]{Bieron2005RaHFS}%
  \BibitemOpen
  \bibfield  {author} {\bibinfo {author} {\bibfnamefont {J.}~\bibnamefont {Biero{\'n}}},\ }\href {https://doi.org/10.1088/0953-4075/38/13/014} {\bibfield  {journal} {\bibinfo  {journal} {J. Phys. B: At. Mol. Opt. Phys.}\ }\textbf {\bibinfo {volume} {38}},\ \bibinfo {pages} {2221} (\bibinfo {year} {2005})}\BibitemShut {NoStop}%
\bibitem [{\citenamefont {Sahoo}\ \emph {et~al.}(2007)\citenamefont {Sahoo}, \citenamefont {Das}, \citenamefont {Chaudhuri}, \citenamefont {Mukherjee}, \citenamefont {Timmermans},\ and\ \citenamefont {Jungmann}}]{Sahoo2007RaPlus}%
  \BibitemOpen
  \bibfield  {author} {\bibinfo {author} {\bibfnamefont {B.~K.}\ \bibnamefont {Sahoo}}, \bibinfo {author} {\bibfnamefont {B.~P.}\ \bibnamefont {Das}}, \bibinfo {author} {\bibfnamefont {R.~K.}\ \bibnamefont {Chaudhuri}}, \bibinfo {author} {\bibfnamefont {D.}~\bibnamefont {Mukherjee}}, \bibinfo {author} {\bibfnamefont {R.~G.~E.}\ \bibnamefont {Timmermans}},\ and\ \bibinfo {author} {\bibfnamefont {K.}~\bibnamefont {Jungmann}},\ }\href {https://doi.org/10.1103/PhysRevA.76.040504} {\bibfield  {journal} {\bibinfo  {journal} {Phys. Rev. A}\ }\textbf {\bibinfo {volume} {76}},\ \bibinfo {pages} {040504(R)} (\bibinfo {year} {2007})}\BibitemShut {NoStop}%
\bibitem [{\citenamefont {Hassouneh}\ and\ \citenamefont {Salah}(2022)}]{HassounehSalah2022RaHFS}%
  \BibitemOpen
  \bibfield  {author} {\bibinfo {author} {\bibfnamefont {O.}~\bibnamefont {Hassouneh}}\ and\ \bibinfo {author} {\bibfnamefont {W.}~\bibnamefont {Salah}},\ }\href {https://doi.org/10.1140/epjp/s13360-022-02422-2} {\bibfield  {journal} {\bibinfo  {journal} {Eur. Phys. J. Plus}\ }\textbf {\bibinfo {volume} {137}},\ \bibinfo {pages} {426} (\bibinfo {year} {2022})}\BibitemShut {NoStop}%
\bibitem [{\citenamefont {Wendt}\ \emph {et~al.}(1987)\citenamefont {Wendt}, \citenamefont {Ahmad}, \citenamefont {Klempt}, \citenamefont {Neugart}, \citenamefont {Otten},\ and\ \citenamefont {Stroke}}]{Wendt1987RaHFS}%
  \BibitemOpen
  \bibfield  {author} {\bibinfo {author} {\bibfnamefont {K.}~\bibnamefont {Wendt}}, \bibinfo {author} {\bibfnamefont {S.~A.}\ \bibnamefont {Ahmad}}, \bibinfo {author} {\bibfnamefont {W.}~\bibnamefont {Klempt}}, \bibinfo {author} {\bibfnamefont {R.}~\bibnamefont {Neugart}}, \bibinfo {author} {\bibfnamefont {E.~W.}\ \bibnamefont {Otten}},\ and\ \bibinfo {author} {\bibfnamefont {H.~H.}\ \bibnamefont {Stroke}},\ }\href {https://doi.org/10.1007/BF01436633} {\bibfield  {journal} {\bibinfo  {journal} {Z. Phys. D At. Mol. Clust.}\ }\textbf {\bibinfo {volume} {4}},\ \bibinfo {pages} {227} (\bibinfo {year} {1987})}\BibitemShut {NoStop}%
\bibitem [{\citenamefont {Neu}\ \emph {et~al.}(1989)\citenamefont {Neu}, \citenamefont {Neugart}, \citenamefont {Otten}, \citenamefont {Passler}, \citenamefont {Wendt}, \citenamefont {Fricke}, \citenamefont {Arnold}, \citenamefont {Kluge},\ and\ \citenamefont {Ulm}}]{Neu1989RaIIQuadrupole}%
  \BibitemOpen
  \bibfield  {author} {\bibinfo {author} {\bibfnamefont {W.}~\bibnamefont {Neu}}, \bibinfo {author} {\bibfnamefont {R.}~\bibnamefont {Neugart}}, \bibinfo {author} {\bibfnamefont {E.-W.}\ \bibnamefont {Otten}}, \bibinfo {author} {\bibfnamefont {G.}~\bibnamefont {Passler}}, \bibinfo {author} {\bibfnamefont {K.}~\bibnamefont {Wendt}}, \bibinfo {author} {\bibfnamefont {B.}~\bibnamefont {Fricke}}, \bibinfo {author} {\bibfnamefont {E.}~\bibnamefont {Arnold}}, \bibinfo {author} {\bibfnamefont {H.~J.}\ \bibnamefont {Kluge}},\ and\ \bibinfo {author} {\bibfnamefont {G.}~\bibnamefont {Ulm}} (\bibinfo {collaboration} {ISOLDE}),\ }\href {https://doi.org/10.1007/BF01444425} {\bibfield  {journal} {\bibinfo  {journal} {Z. Phys. D At. Mol. Clust.}\ }\textbf {\bibinfo {volume} {11}},\ \bibinfo {pages} {105} (\bibinfo {year} {1989})}\BibitemShut {NoStop}%
\bibitem [{\citenamefont {Lynch}\ \emph {et~al.}(2018)\citenamefont {Lynch}, \citenamefont {Wilkins}, \citenamefont {Billowes}, \citenamefont {Binnersley}, \citenamefont {Bissell}, \citenamefont {Chrysalidis}, \citenamefont {Cocolios}, \citenamefont {Day~Goodacre}, \citenamefont {de~Groote}, \citenamefont {Farooq-Smith}, \citenamefont {Fedorov}, \citenamefont {Fedosseev}, \citenamefont {Flanagan}, \citenamefont {Franchoo}, \citenamefont {Garcia~Ruiz}, \citenamefont {Gins}, \citenamefont {Heinke}, \citenamefont {Koszor{\'u}s}, \citenamefont {Marsh}, \citenamefont {Molkanov}, \citenamefont {Naubereit}, \citenamefont {Neyens}, \citenamefont {Ricketts}, \citenamefont {Rothe}, \citenamefont {Seiffert}, \citenamefont {Seliverstov}, \citenamefont {Stroke}, \citenamefont {Studer}, \citenamefont {Vernon}, \citenamefont {Wendt},\ and\ \citenamefont {Yang}}]{Lynch2018RaCRIS}%
  \BibitemOpen
  \bibfield  {author} {\bibinfo {author} {\bibfnamefont {K.~M.}\ \bibnamefont {Lynch}}, \bibinfo {author} {\bibfnamefont {S.~G.}\ \bibnamefont {Wilkins}}, \bibinfo {author} {\bibfnamefont {J.}~\bibnamefont {Billowes}}, \bibinfo {author} {\bibfnamefont {C.~L.}\ \bibnamefont {Binnersley}}, \bibinfo {author} {\bibfnamefont {M.~L.}\ \bibnamefont {Bissell}}, \bibinfo {author} {\bibfnamefont {K.}~\bibnamefont {Chrysalidis}}, \bibinfo {author} {\bibfnamefont {T.~E.}\ \bibnamefont {Cocolios}}, \bibinfo {author} {\bibfnamefont {T.}~\bibnamefont {Day~Goodacre}}, \bibinfo {author} {\bibfnamefont {R.~P.}\ \bibnamefont {de~Groote}}, \bibinfo {author} {\bibfnamefont {G.~J.}\ \bibnamefont {Farooq-Smith}}, \bibinfo {author} {\bibfnamefont {D.~V.}\ \bibnamefont {Fedorov}}, \bibinfo {author} {\bibfnamefont {V.~N.}\ \bibnamefont {Fedosseev}}, \bibinfo {author} {\bibfnamefont {K.~T.}\ \bibnamefont {Flanagan}}, \bibinfo {author} {\bibfnamefont {S.}~\bibnamefont {Franchoo}}, \bibinfo {author} {\bibfnamefont {R.~F.}\ \bibnamefont
  {Garcia~Ruiz}}, \bibinfo {author} {\bibfnamefont {W.}~\bibnamefont {Gins}}, \bibinfo {author} {\bibfnamefont {R.}~\bibnamefont {Heinke}}, \bibinfo {author} {\bibfnamefont {{\'A}.}~\bibnamefont {Koszor{\'u}s}}, \bibinfo {author} {\bibfnamefont {B.~A.}\ \bibnamefont {Marsh}}, \bibinfo {author} {\bibfnamefont {P.~L.}\ \bibnamefont {Molkanov}}, \bibinfo {author} {\bibfnamefont {P.}~\bibnamefont {Naubereit}}, \bibinfo {author} {\bibfnamefont {G.}~\bibnamefont {Neyens}}, \bibinfo {author} {\bibfnamefont {C.~M.}\ \bibnamefont {Ricketts}}, \bibinfo {author} {\bibfnamefont {S.}~\bibnamefont {Rothe}}, \bibinfo {author} {\bibfnamefont {C.}~\bibnamefont {Seiffert}}, \bibinfo {author} {\bibfnamefont {M.~D.}\ \bibnamefont {Seliverstov}}, \bibinfo {author} {\bibfnamefont {H.~H.}\ \bibnamefont {Stroke}}, \bibinfo {author} {\bibfnamefont {D.}~\bibnamefont {Studer}}, \bibinfo {author} {\bibfnamefont {A.~R.}\ \bibnamefont {Vernon}}, \bibinfo {author} {\bibfnamefont {K.~D.~A.}\ \bibnamefont {Wendt}},\ and\ \bibinfo {author}
  {\bibfnamefont {X.~F.}\ \bibnamefont {Yang}},\ }\href {https://doi.org/10.1103/PhysRevC.97.024309} {\bibfield  {journal} {\bibinfo  {journal} {Phys. Rev. C}\ }\textbf {\bibinfo {volume} {97}},\ \bibinfo {pages} {024309} (\bibinfo {year} {2018})}\BibitemShut {NoStop}%
\end{thebibliography}

%

\end{document}